\documentclass{article} 
\usepackage{iclr2027_conference,times}
\usepackage{graphicx}
\usepackage{placeins}
\usepackage{amsmath,amsfonts,bm}
\usepackage{xcolor}
\usepackage{tcolorbox}
\usepackage{enumitem}
\tcbuselibrary{breakable,skins}
\definecolor{benchmarkblue}{HTML}{52768E}
\definecolor{benchmarkpaper}{HTML}{F5F7F9}
\newtcolorbox[auto counter]{benchmarkprompt}[2]{enhanced,breakable,label={#2},
  colback=benchmarkpaper,colframe=benchmarkblue,
  colbacktitle=benchmarkblue!10,coltitle=black,
  title={Prompt~\thetcbcounter. #1},
  title after break={Prompt~\thetcbcounter. #1 (continued)},
  fonttitle=\small\bfseries,fontupper=\small\raggedright,
  boxrule=0.45pt,arc=1.2mm,left=7pt,right=7pt,top=5pt,bottom=5pt,
  before skip=9pt,after skip=9pt,pad at break*=2mm}

\usepackage{hyperref}
\usepackage{url}
\usepackage{tabularx}
\usepackage{booktabs}
\usepackage{array}
\usepackage{xcolor}
\usepackage{colortbl}
\usepackage{pifont}
\usepackage{multirow}


\title{Edit Who Speaks, Control How They Speak: Global Timbre Editing and Local Instruction Control for TTS}

\author{
Junchuan Zhao$^{1}$,
Chenglin Xu$^{2}$\thanks{Corresponding authors.} ,
Wei Zeng$^{1}$,
Haoyang Li$^{3}$,
Yiwen Guo$^{4*}$,
Ye Wang$^{1}$ \\
$^{1}$National University of Singapore \\
$^{2}$LIGHTSPEED \\
$^{3}$Nanyang Technological University \\
$^{4}$Independent Researcher
}

\iclrfinalcopy

\begin{document}

\maketitle

\begin{abstract}
Instruction-based text-to-speech (TTS) offers control over voice characteristics and speech expression through interfaces including voice cloning and text-based voice design. Voice cloning reproduces a reference voice, whereas text-based voice design creates a voice from a natural-language description. However, neither interface directly enables users to modify the timbre of a given reference and synthesize speech with the modified voice. Meanwhile, utterance-level expressive instructions leave changes across individual text segments underspecified. We introduce \textbf{EDICT}, a framework that unifies global timbre editing and local expressive control by using an edited acoustic reference to anchor voice identity across segments. To enable synthesis with an instruction-edited voice, EDICT combines reference audio with structured timbre edits to generate an edited reference in codec-token space. This representation serves as a shared voice anchor for a frozen TTS backbone, allowing segment-specific natural-language instructions to guide expression. To accommodate instruction changes while supporting acoustic continuity, EDICT rebuilds the KV cache at each segment boundary, refreshing instruction conditioning while retaining bounded acoustic context from previously generated speech. Evaluations on our proposed TimbreEdit-Bench and IntraTTS-Bench demonstrate improved timbre editing and a favorable balance between local instruction adherence, speaker consistency, and transition quality. Audio demos are available\footnote{\url{https://danny-nus.github.io/EDICT-demopage/}}.
\end{abstract}

\section{Introduction}
\label{sec:introduction}
Instruction-based TTS not only synthesizes speech from the given text but also allows users to explicitly control voice characteristics and speech delivery
\citep{DBLP:conf/icassp/GuoLWZT23,DBLP:journals/taslp/YangLHWM24,zhang2026poly}.
Narration and dialog often require control at two temporal scopes: voice characteristics should remain consistent throughout an utterance, while speech delivery may vary across individual text segments. For example, a brighter version of a reference voice may serve as a shared voice anchor throughout an utterance, while the delivery changes from calm to excited and finally to soft. We study TTS requests that combine
\emph{global} editing of a reference voice with \emph{local} instructions
for successive segments of one utterance.

At the global level, existing approaches specify a target voice either through an acoustic reference, as in voice cloning, or through a textual description, as in voice design
\citep{wang2023neural,DBLP:conf/iclr/LengGSJ0LLYZS0024,hu2026qwen3,lee2026fc,zhao2025prosody}.
Reference-conditioned editing provides a complementary interface: users specify changes relative to an existing voice, while unspecified attributes are intended to remain unchanged.
Based on text-guided voice conversion and editing
\citep{DBLP:conf/interspeech/HaiTWQE24,anastassiou2024voiceshop,huang2024instructspeech},
we use the resulting edited voice as a shared conditioning reference for synthesizing new text under changing delivery instructions, thereby integrating timbre editing into the TTS pipeline.

At the local level, segment-specific instructions must align with
their text spans while preserving the globally edited voice characteristics and maintaining natural
transitions. Most instruction-based TTS systems use a single prompt for the entire utterance, while only a few support the association of individual text spans with their own instructions within an utterance. TED-TTS establishes segment-aware emotion
and duration control, but its masking mechanism does not refresh
the acoustic history, allowing cached key--value (KV)
states to carry preceding conditions into later segments and cause
\emph{control leakage}
\citep{DBLP:conf/slt/WuWETTTLXZLK24,jeong2025tts,lemerle2025linastyle,DBLP:conf/acl/LiangLWLZW26}.
Retaining the full history may weaken new instructions, whereas
discarding it removes the context needed for natural continuation.
Kang et al.\ mitigate \emph{style self-referencing} through
KV-cache swapping and sliding-window attention, but focus on a single
source-to-target transition and report a trade-off between transition
strength and speaker similarity \citep{kang2026unlocking}.
EDICT rebuilds the cache at every boundary from the current
instruction, edited reference, and bounded acoustic context, enabling
repeated instruction switching while preserving identity and natural
transitions.


We introduce \textbf{EDICT}, a controllable TTS framework whose trainable timbre editor maps a global request to attribute operations and edits the source voice, producing a codec reference shared across all segments of a frozen TTS backbone.
To mitigate control leakage, EDICT reconstructs the KV cache at instruction boundaries by recomputing selected initial and recent acoustic states conditioned on the next text, instruction, and shared reference.
Initial context provides voice cues, while recent context supports continuity.
We propose TimbreEdit-Bench and IntraTTS-Bench to evaluate, respectively, requested voice edits and preservation of unspecified attributes, and local instruction following with consistent speaker identity and natural transitions.
Experiments on Qwen3-TTS-VoiceDesign (Qwen3-TTS-VD) and CosyVoice~2 assess KV cache reconstruction across backbones and its combination with timbre editing, showing improved instruction following over the respective baselines for timbre editing and intra-utterance control.

In summary, this paper makes the following contributions:
\begin{itemize}
    \item \textbf{Reference-conditioned timbre editing for TTS.}
    We develop a timbre editor that modifies a source voice through
    relative instructions, producing a reusable codec reference for
    new texts and delivery conditions.

    \item \textbf{Segment-level control through KV cache reconstruction.}
    We reconstruct the retained acoustic context under new instructions
    to mitigate control leakage and balance local adherence with
    continuity, without additional backbone training.

    \item \textbf{Bilingual benchmarks and compositional evaluation.}
    We introduce TimbreEdit-Bench and IntraTTS-Bench to evaluate global timbre editing, local instruction control, and their composition
    through objective metrics, listening tests, and ablations.
\end{itemize}

\section{Related Work}
\label{sec:related_work}
\paragraph{Language-based voice and style control.}
Early work represented speaking style through learned latent tokens
\citep{DBLP:conf/icml/WangSZRBSXJRS18}.
PromptTTS, InstructTTS, PromptStyle, and TextrolSpeech control voice
and delivery through descriptions
\citep{DBLP:conf/icassp/GuoLWZT23,DBLP:journals/taslp/YangLHWM24,DBLP:conf/interspeech/LiuZLCW0L23,DBLP:conf/icassp/JiZ00CDHZ24}.
More recent systems use language models to condition speech generation
on voice and delivery instructions
\citep{DBLP:conf/mm/0002QJZLZ0024,du2025cosyvoice,hu2026qwen3,DBLP:conf/mm/YangY00C0W0NL0D25,zhang2026poly}.
ControlSpeech and FlexiVoice combine voice references with language-based
style control; ReStyle-TTS adjusts style relative to a reference, and
Vevo separates timbre and style prompts
\citep{DBLP:conf/acl/JiCWZ0000CZ025,chen2026flexivoice,DBLP:conf/acl/LiJLGHC26,DBLP:conf/iclr/ZhangZPTMLHLWCH25}.

\paragraph{Reference-anchored voice editing.}
Voice cloning specifies voices acoustically
\citep{wang2023neural,DBLP:conf/nips/LeVSKSMWMAMH23,du2024cosyvoice},
while voice design specifies them textually
\citep{DBLP:conf/iclr/LengGSJ0LLYZS0024,DBLP:conf/icassp/ShimizuYKSDKT24,hu2026voicesculptor}.
Voice editing applies instructed changes to a source voice
\citep{huang2024instructspeech}.
VoxEditor represents relative attribute edits \citep{sheng2025voice};
DreamVoice and TES-VC use text-guided conversion
\citep{DBLP:conf/interspeech/HaiTWQE24,DBLP:conf/interspeech/JinYZW25};
VoiceShop supports selective voice changes \citep{anastassiou2024voiceshop}.
Recent systems integrate editing with TTS or unify generation and
editing \citep{pei2026unified,hai2026voicedesigner,wang2026dots}.
EDICT produces a reusable edited codec reference for subsequent TTS
under changing local instructions. We evaluate requested changes and
preservation separately, since target-voice matching does not establish
preservation of unspecified attributes.

\paragraph{Local control and acoustic history.}
Time-varying style and local prosody
\citep{DBLP:conf/icassp/ChenR22,DBLP:conf/ssw/LengletPB23,DBLP:conf/slt/WuWETTTLXZLK24,jeong2025tts} have expanded
to word-level styles, emotion trajectories, and timing control
\citep{DBLP:conf/nips/WangWGGQMHYWCCW25,liu2026emotra,zhou2026sequential,mai2026magic,lemerle2025linastyle}.
For training-free control, TED-TTS uses segment-aware causal masking
and monotonic alignment \citep{DBLP:conf/acl/LiangLWLZW26}.
Kang et al.\ address acoustic \emph{style self-referencing} through
KV-cache swapping and sliding-window attention for a source-to-target
transition \citep{kang2026unlocking}.
EDICT recomputes retained acoustic states at successive instruction
switches and uses an edited reference throughout the utterance.
Appendix~\ref{app:method_comparison} defines the tasks and comparison scope.

\section{Methodology}
\label{sec:method}
\begin{figure}[!t]
    \centering
    \includegraphics[width=\linewidth]{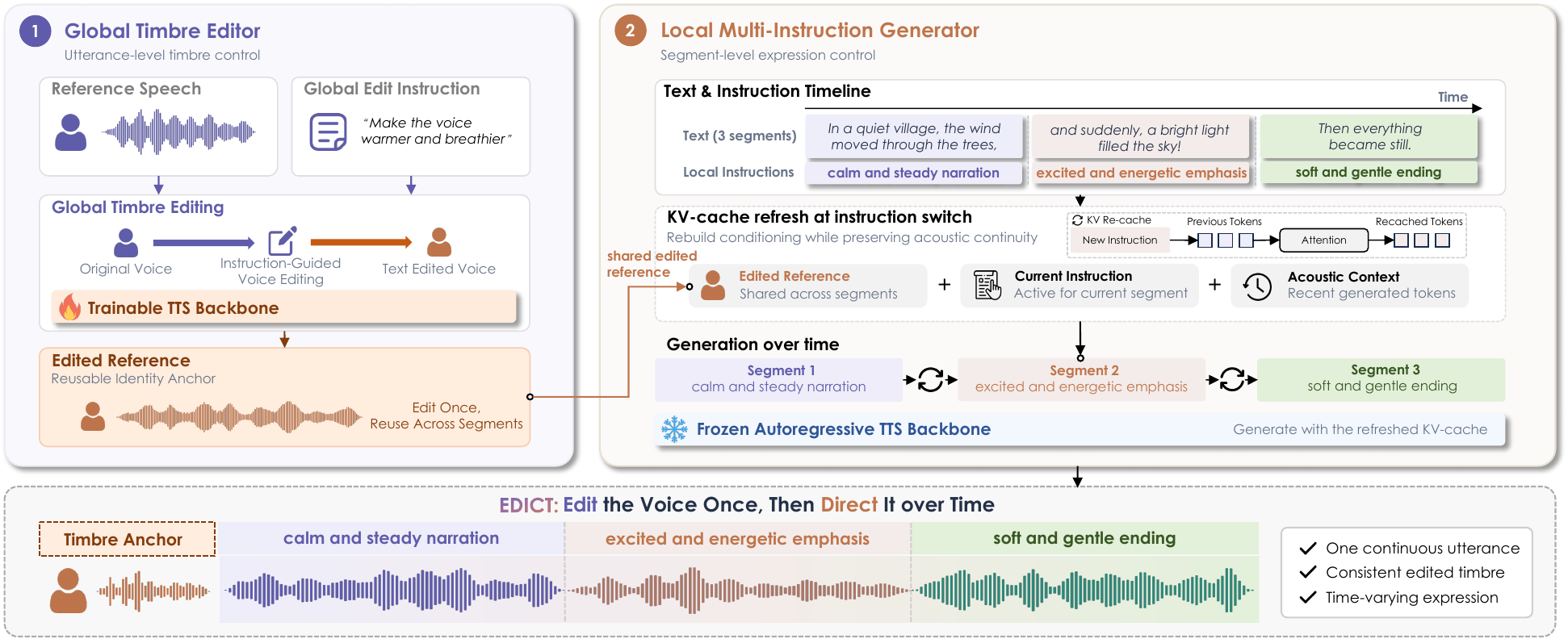}
    \caption{\textbf{EDICT combines global voice editing with local delivery control.}
    The trainable editor uses the source voice and global edit instruction
    to generate an edited codec reference. This reference conditions
    every segment of the frozen TTS backbone, whose KV cache is rebuilt
    at instruction boundaries using selected acoustic context.
    The segments form one continuous utterance.}
    \label{fig:edict_overview}
\end{figure}

\subsection{Task Formulation and Overview}
\label{sec:method_overview}
At inference, an LLM parses the global request into attribute operations, the timbre editor produces an edited codec reference shared across segments, and a frozen TTS backbone generates the segments sequentially while rebuilding its KV cache at local-instruction boundaries. Specifically, given a reference codec sequence $\mathbf{c}_{\mathrm{ref}}$,
a timbre-edit request $\mathcal{I}_{\mathrm{edit}}$, and
$N$ text--instruction pairs
$\{(\mathbf{x}_k,\mathcal{I}_k)\}_{k=1}^{N}$,
EDICT generates successive segments in a shared edited voice.
As shown in Figure~\ref{fig:edict_overview}, a learned editor
$p_\theta$ and a frozen TTS model $p_\phi$ form a two-stage cascade:
\begin{equation}
\begin{aligned}
\mathbf{c}'_{\mathrm{ref}}
\sim p_{\theta}\!\left(\cdot\mid
\mathbf{c}_{\mathrm{ref}},\mathbf{z}_{\mathrm{edit}},\mathbf{x}_{\mathrm{edit}}\right),
\quad
\mathbf{c}_{1:N}
\sim p_{\phi}\!\left(\cdot\mid
\mathbf{c}'_{\mathrm{ref}},\{(\mathbf{x}_k,\mathcal{I}_k)\}_{k=1}^{N}\right),
\end{aligned}
\label{eq:edict_cascade}
\end{equation}
where $\mathbf{z}_{\mathrm{edit}}$ encodes the requested attribute
operations and $\mathbf{x}_{\mathrm{edit}}$ specifies the text spoken
in the edited reference. The edited codec reference conditions every
segment, while local instructions change their delivery.
Segment-boundary reconstruction produces one continuous codec stream
for waveform decoding. 

\subsection{Reference-Conditioned Timbre Editing}
\label{sec:method_timbre_editing}

\paragraph{Attribute-based instruction encoding.}
Following work on relative voice editing and perceptual voice-quality
control \citep{sheng2025voice,rautenberg2025speech}, we encode edits
along nine attributes: perceived gender, age impression, pitch register,
brightness, vocal weight, resonance focus, roughness, breathiness, and
nasality.
Gender and age are categorical; the others use three ordered levels,
with definitions and values in Table~\ref{tab:timbre_attributes}
(Appendix~\ref{app:training_data}).
At inference, an LLM maps structured or descriptive
$\mathcal{I}_{\mathrm{edit}}$ to operations specifying direction and
magnitude for ordinal changes, target categories for categorical changes,
and \texttt{<keep>} for unmentioned attributes.
For example, \texttt{<brightness\_up\_1>} denotes low-to-mid or mid-to-high,
while \texttt{<brightness\_up\_2>} denotes low-to-high.
Operations are serialized in a fixed attribute order into
$\mathbf{z}_{\mathrm{edit}}$ and interpreted together with the acoustic
reference.
During training, profile differences determine the operations.

\paragraph{Paired supervision.}
Qwen3-14B~\citep{yang2025qwen3} describes source and target profiles,
and Qwen3-TTS-VD~\citep{hu2026qwen3} synthesizes their
Chinese and English speech under a shared neutral-delivery requirement.
The requested edits and keep operations are derived from differences between the source and target profiles.
Training tuples
$(\mathbf{c}_{\mathrm{ref}},\mathbf{z}_{\mathrm{edit}},
\mathbf{x}_{\mathrm{edit}},\mathbf{c}_{\mathrm{tgt}})$
cover single-attribute and compositional edits with same or different
texts, within and across languages. The target transcript specifies
the edited reference's content, allowing it to differ from the source.
Appendix~\ref{app:training_data} details construction and filtering.

\paragraph{Training.}
We initialize the editor from Qwen3-TTS-Base~\citep{hu2026qwen3}
and perform supervised fine-tuning (SFT) to predict target codecs
conditioned on
$g=(\mathbf c_{\mathrm{ref}},\mathbf z_{\mathrm{edit}},
\mathbf x_{\mathrm{edit}})$,
using teacher-forced cross-entropy over all codebooks at
target audio positions.
We then apply reward-weighted DPO~\citep{rafailov2023direct}
with auxiliary SFT, denoted wDPO.
Candidates sampled from the frozen SFT model $p_{\mathrm{ref}}$
are scored by WavLM speaker-embedding cosine similarity to
the paired target~\citep{chen2022wavlm}.
These scores determine preference pairs
$(\mathbf c_g^+,\mathbf c_{g,j}^-)$ and positive
reward-gap weights $w_{g,j}$.
Defining $q_\theta(\mathbf c\mid g)
=\log[p_\theta(\mathbf c\mid g)/p_{\mathrm{ref}}(\mathbf c\mid g)]$,
we minimize: 
\begin{equation}
\mathcal L_{\mathrm{wDPO}}
=
-\frac{
    \sum_{(g,j)\in\mathcal B} w_{g,j}
    \log\sigma\!\left(
        \beta\left[
        q_\theta(\mathbf c_g^+\mid g)
        -q_\theta(\mathbf c_{g,j}^-\mid g)
        \right]\right)
}{
    \sum_{(g,j)\in\mathcal B} w_{g,j}
}
+\lambda_{\mathrm{SFT}}\mathcal L_{\mathrm{SFT}},
\label{eq:timbre_weighted_dpo}
\end{equation}
where $\mathcal B$ is a minibatch of preference pairs,
$\sigma$ is the sigmoid, $\beta$ scales the preference logit,
and $\mathcal L_{\mathrm{SFT}}$ is the codec cross-entropy
on chosen candidates.
Preference pairs and weights remain fixed during post-training.
Appendix~\ref{app:implementation_training} details candidate
sampling, pair selection, and training settings.

\begin{figure}[!t]
    \centering
    \includegraphics[width=\linewidth]{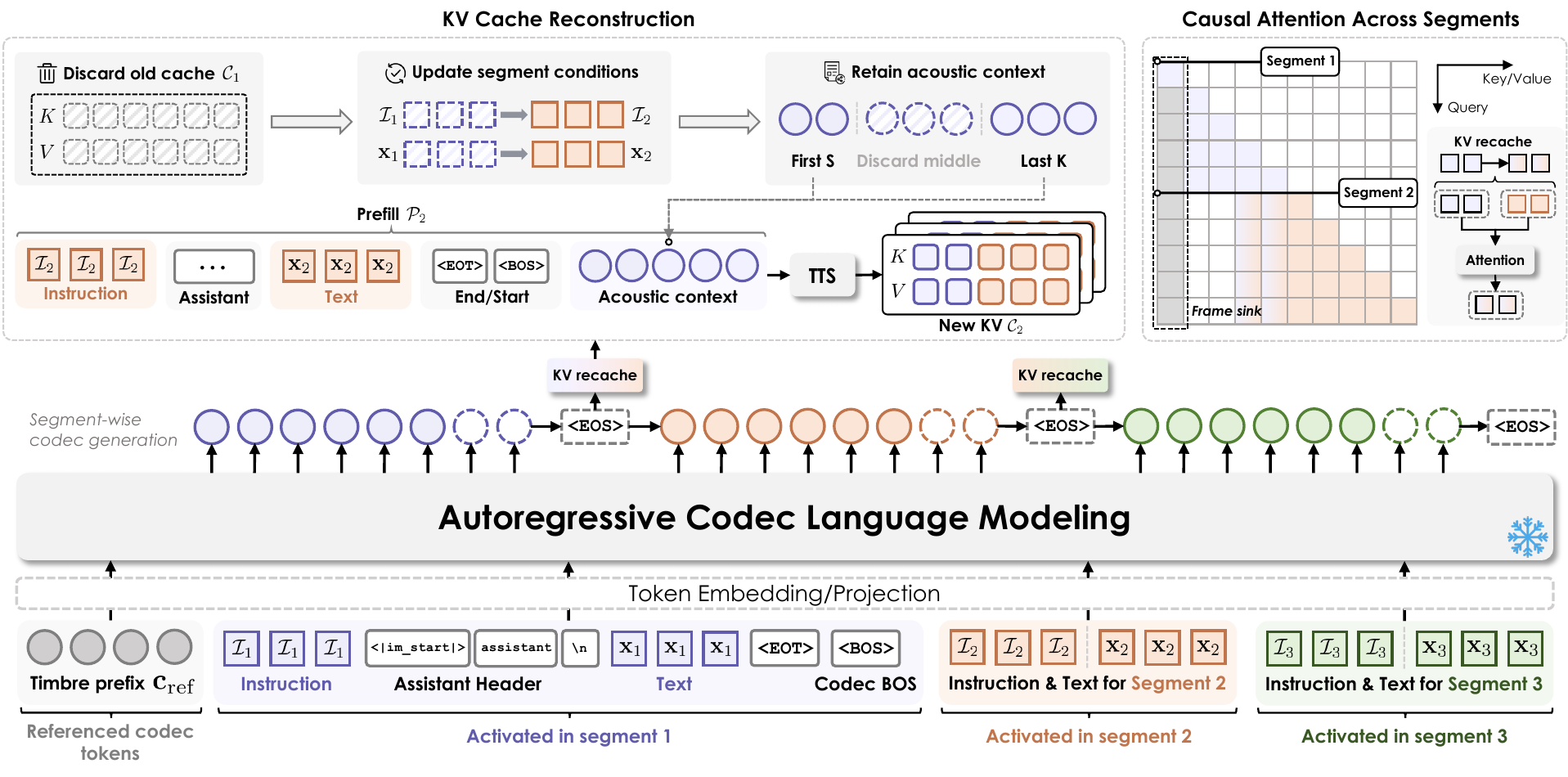}
    \caption{\textbf{Segment-boundary KV cache reconstruction.}
    Each non-final \texttt{<EOS>} triggers a fresh prefill containing the
    next instruction and text, the shared edited reference, and acoustic
    inputs from the first $S$ and most recent $K$ generated positions.
    The upper-left panel shows replacement of the old cache; the
    upper-right panel shows causal attention after reconstruction.
    The lower timeline illustrates successive instruction switches
    within one codec stream.}
    \label{fig:kv_recache}
\end{figure}

\subsection{Segment-Level Instruction Control}
\label{sec:method_segment_control}
The edited reference $\mathbf{c}'_{\mathrm{ref}}$ remains fixed across
segments, while $\mathcal{I}_k$ controls each segment's emotion,
speaking rate, or prosody. This stage uses a frozen autoregressive TTS backbone and requires
no additional training. Figure~\ref{fig:kv_recache} shows how
generation proceeds across instruction boundaries.
Segment-aware masks such as those in TED-TTS \citep{DBLP:conf/acl/LiangLWLZW26} restrict direct access
to other segments' conditions,
but historical acoustic states may still encode those conditions.
Drawing on LongLive's recaching strategy \citep{yang2025longlive},
we reconstruct these states under the next instruction.

\paragraph{KV cache reconstruction.}
A non-final \texttt{<EOS>} triggers replacement of
$(\mathcal{I}_k,\mathbf{x}_k)$ with
$(\mathcal{I}_{k+1},\mathbf{x}_{k+1})$.
We assemble a fresh prefill $\mathcal{P}_{k+1}$ containing the new
instruction, shared edited reference, new text, and native template
and boundary tokens. We then append acoustic input embeddings for
the first $S$ and most recent $K$ generated codec positions.
The old KV cache is discarded, and a causal forward pass through
the frozen model reconstructs the entire prefill cache:
\begin{equation}
\mathcal{C}_{k+1}
=\operatorname{Prefill}_{\phi}(\mathcal{P}_{k+1})
=\left\{\left(
\mathbf{K}_{\mathcal{P}_{k+1}}^{(\ell)},
\mathbf{V}_{\mathcal{P}_{k+1}}^{(\ell)}
\right)\right\}_{\ell=1}^{L},
\label{eq:kv_reconstruction}
\end{equation}
where $L$ is the number of Transformer layers.
Acoustic input embeddings are reused, but their keys and values are
recomputed under the updated conditions. Initial context retains
evidence from the utterance's start; recent context supplies immediate
continuation history. Reconstruction bounds the retained history,
while the cache grows normally within each segment.
Section~\ref{sec:results_ablations} tests the role of each context
source and compares reconstruction with reuse of the old KV states.

\paragraph{Continuous decoding.}
For a single attention head, let $j$ index an acoustic input
position processed after the reconstructed prefill.
The query $\mathbf{q}_{k+1,j}$ attends to the rebuilt prefill
and the current segment's acoustic inputs through position $j$
(layer indices omitted):
\begin{equation}
\operatorname{Attn}(\mathbf{q}_{k+1,j})
=
\operatorname{softmax}\!\left(
\frac{
\mathbf{q}_{k+1,j}
[\mathbf{K}_{\mathcal{P}_{k+1}};
 \mathbf{K}_{k+1,\leq j}]^\top
}{\sqrt{d_h}}
\right)
[\mathbf{V}_{\mathcal{P}_{k+1}};
 \mathbf{V}_{k+1,\leq j}],
\label{eq:segment_attention}
\end{equation}
where $[\,;\,]$ denotes concatenation along the sequence
dimension, $d_h$ is the head dimension, and
$(\mathbf{K}_{k+1,\leq j},\mathbf{V}_{k+1,\leq j})$
contains the current segment's KV states through input
position $j$. These positions are already observed inputs;
the attention output is used to predict the next codec frame.
History omitted from the cache remains in the output codec stream.
Non-final \texttt{<EOS>} markers trigger reconstruction without entering
the acoustic output; the final \texttt{<EOS>} ends generation.
All segments are decoded into one waveform.
Appendix~\ref{app:implementation_decoding} details decoding, and
Appendix~\ref{app:attention_analysis} visualizes attention patterns.

\section{Experiments}
\label{sec:experiments}
\begin{figure}[!t]
    \centering
    \includegraphics[width=\linewidth]{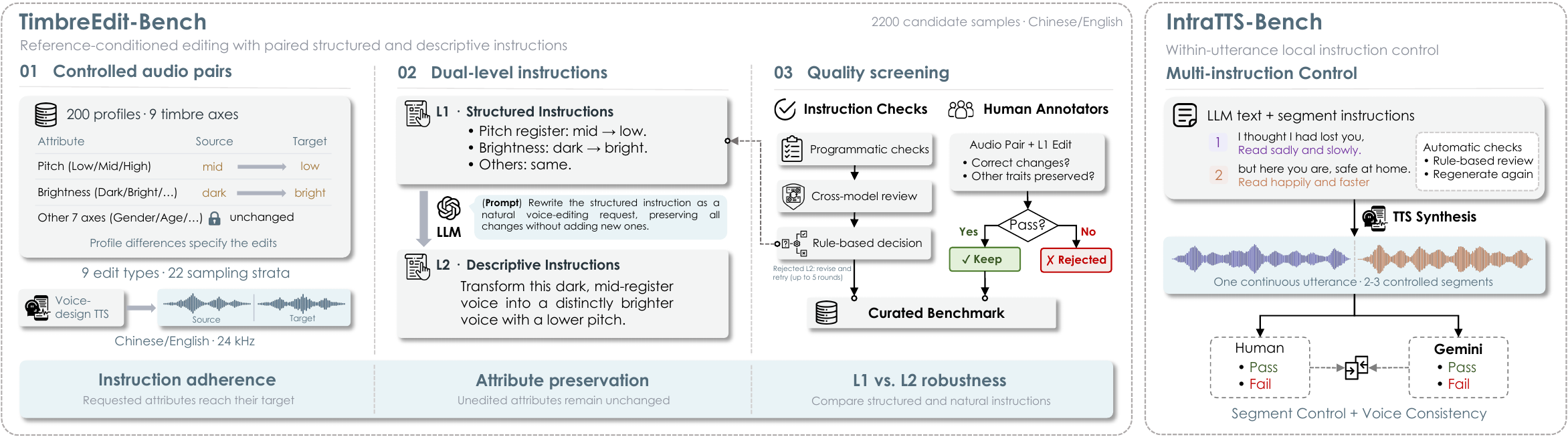}
    \caption{\textbf{Benchmark construction and evaluation.}
Left: TimbreEdit-Bench pairs source and target voices with editing
instructions. Right: IntraTTS-Bench aligns delivery instructions
with successive text segments.}
    \label{fig:benchmarks}
\end{figure}

\subsection{Data and Benchmarks}
\label{sec:experiments_datasets}
\paragraph{Training data.}
We construct 1M directed editing pairs from a synthetic Chinese--English
speech bank generated by Qwen3-TTS-VD using nine-attribute
timbre profiles. We use 5,800 profiles for training and hold out 200,
excluding held-out profiles from both sides of every training pair.
Edit operations are derived from source--target profile differences,
covering single-attribute and compositional edits within and across
languages. Offline post-training conditions are sampled from the same
training pairs. Appendix~\ref{app:training_data} details the construction
and data distributions.

\paragraph{Evaluation benchmarks.}
\textbf{TimbreEdit-Bench} contains 1,955 source--target pairs from
200 held-out profiles across nine edit categories.
Three annotators screen for audible edits, attribute preservation,
and audio quality; retained pairs have semantically matched structured
and descriptive instructions.
\textbf{IntraTTS-Bench} contains 997 requests with two or three
text segments and corresponding delivery instructions to evaluate
local control and continuity.
Both benchmarks cover Chinese and English.
Figure~\ref{fig:benchmarks} summarizes the workflow, and
Appendix~\ref{app:benchmark_construction} details construction and screening.

\subsection{Implementation Details}
\label{sec:experiments_implementation}
We initialize the editor from Qwen3-TTS-12Hz-1.7B-Base and fine-tune
all parameters, including codec predictors.
SFT is followed by one round of wDPO using preference pairs
constructed from 20K training conditions, with $M=16$ sampled
candidates per condition.
We set the DPO scale $\beta=0.1$ and auxiliary SFT coefficient
$\lambda_{\mathrm{SFT}}=0.01$.
Qwen3-14B parses editing requests.
Local generation uses frozen Qwen3-TTS-VD,
with CosyVoice~2 as a second backbone and default context
$(S,K)=(4,5)$.
For local control, \textbf{Concat.} synthesizes segments independently,
\textbf{Joint} prompts the complete utterance in one pass, and our
\textbf{TED-TTS} adaptation switches segment conditions through
attention masking and monotonic alignment without KV cache
reconstruction.
Appendix~\ref{app:implementation_details} provides the remaining
training settings and baseline and ablation configurations.

\subsection{Evaluation Metrics}
\label{sec:experiments_metrics}
Chinese \textbf{CER} and English \textbf{WER} assess intelligibility;
\textbf{UTMOS}~\citep{saeki2022utmos} and \textbf{DNSMOS} OVRL
\citep{reddy2022dnsmos} estimate perceptual quality.
WavLM~\citep{chen2022wavlm} and ERes2Net~\citep{chen2023enhanced}
measure target/source similarity (\textbf{SIM-T/R}); WavLM also
measures inter-segment similarity (\textbf{SIM-I}).
Following InstructTTSEval~\citep{huang2025instructttseval},
Gemini 3.1 Pro evaluates instruction compliance (\textbf{SegAcc}),
requested emotion (\textbf{EmoAcc}), transitions (\textbf{Trans}),
and attribute-level editing (\textbf{Up/Down}, \textbf{Category},
and \textbf{Preservation}).
Human evaluation uses five-point ratings of naturalness
(\textbf{N-MOS}), edit accuracy
(\textbf{Edit-MOS}), target similarity (\textbf{S-MOS}), instruction
following (\textbf{Instr-MOS}), and transitions (\textbf{Trans-MOS}),
with 95\% confidence intervals from audio-item bootstrapping.
Appendix~\ref{app:implementation_details} provides training settings,  baseline and ablation configurations.

\section{Results}
\label{sec:results}
\begin{table}[!t]
\centering
\caption{\textbf{Objective results on TimbreEdit-Bench.}
WER/CER reports English WER / Chinese CER in percent.
SIM-T/R denotes target/source similarity;
$\Delta=\mathrm{SIM\text{-}T}-\mathrm{SIM\text{-}R}$.
Best and second-best scores within each instruction form are
\textbf{bold} and \underline{underlined}, with CER and WER ranked separately;
SIM-R is unranked.
Target-reference voice conversion receives target audio and is evaluated
separately on a same-text subset, with its best similarity and $\Delta$
scores in \textbf{bold}.}
\label{tab:res_timbre_main}
\footnotesize
\setlength{\tabcolsep}{2.5pt}
\renewcommand{\arraystretch}{1.24}
\resizebox{0.95\linewidth}{!}{%
\begin{tabular}{@{}ll*{9}{c}@{}}
\toprule
\multirow{2}{*}{\textbf{System}}
 & \multirow{2}{*}{\textbf{Instruction}}
 & \multicolumn{3}{c}{\textbf{Content and quality}}
 & \multicolumn{3}{c}{\textbf{WavLM}}
 & \multicolumn{3}{c}{\textbf{ERes2Net}} \\
\cmidrule(lr){3-5}\cmidrule(lr){6-8}\cmidrule(l){9-11}
 & & WER/CER (\%)$\downarrow$
 & UTMOS$\uparrow$ & DNSMOS$\uparrow$
 & SIM-T$\uparrow$ & SIM-R & $\Delta\uparrow$
 & SIM-T$\uparrow$ & SIM-R & $\Delta\uparrow$ \\
\midrule
\multicolumn{11}{@{}l}{\textit{Instruction-conditioned systems}} \\
\addlinespace[3pt]
\multirow{2}{*}{Qwen3-TTS-VD}
 & Structured
 & \underline{4.5} / \textbf{0.3}
 & \underline{3.861} & \underline{3.946}
 & \underline{0.701} & 0.871 & $\underline{-0.170}$
 & \underline{0.450} & 0.745 & $\underline{-0.295}$ \\
 & Descriptive
 & \textbf{3.0} / \textbf{0.3}
 & \underline{3.830} & \underline{3.965}
 & \underline{0.676} & 0.832 & $\underline{-0.156}$
 & \underline{0.435} & 0.714 & $\underline{-0.279}$ \\
\addlinespace[4pt]
\multirow{2}{*}{Qwen3-TTS-Base}
 & Structured
 & 5.1 / 1.9 & 3.742 & 3.805
 & 0.632 & 0.950 & $-0.318$
 & 0.431 & 0.845 & $-0.414$ \\
 & Descriptive
 & \underline{3.3} / \underline{1.1} & 3.756 & 3.776
 & 0.637 & 0.950 & $-0.313$
 & 0.434 & 0.849 & $-0.415$ \\
\addlinespace[4pt]
\multirow{2}{*}{CosyVoice 2}
 & Structured
 & 6.9 / 4.2 & \textbf{3.941} & \textbf{4.065}
 & 0.646 & 0.918 & $-0.272$
 & 0.396 & 0.825 & $-0.429$ \\
 & Descriptive
 & 7.9 / 2.5 & \textbf{4.021} & \textbf{4.104}
 & 0.644 & 0.937 & $-0.293$
 & 0.419 & 0.854 & $-0.435$ \\
\addlinespace[4pt]
\multirow{2}{*}{\textbf{EDICT-Stage I}}
 & Structured
 & \textbf{4.1} / \underline{1.7} & 3.830 & 3.942
 & \textbf{0.819} & 0.620 & \textbf{0.199}
 & \textbf{0.653} & 0.442 & \textbf{0.211} \\
 & Descriptive
 & 4.0 / \underline{1.1} & 3.816 & 3.963
 & \textbf{0.793} & 0.598 & \textbf{0.195}
 & \textbf{0.648} & 0.441 & \textbf{0.207} \\
\midrule
\multicolumn{11}{@{}l}{\textit{Target-reference voice conversion}} \\
\addlinespace[3pt]
FreeVC & N/A
 & 3.2 / 2.0 & 3.911 & 3.968
 & 0.873 & 0.671 & 0.202
 & 0.570 & 0.387 & 0.183 \\
Seed-VC & N/A
 & 3.0 / 2.5 & 3.342 & 3.893
 & \textbf{0.957} & 0.659 & \textbf{0.298}
 & \textbf{0.889} & 0.463 & \textbf{0.427} \\
\bottomrule
\end{tabular}%
}
\vspace{-12pt}
\end{table}

\subsection{Timbre Editing}
\label{sec:results_timbre_editing}

Table~\ref{tab:res_timbre_main} shows that \textit{EDICT-Stage I}
(Section~\ref{sec:method_timbre_editing}) achieves the highest SIM-T
among instruction-conditioned systems under both instruction forms,
with gains on both ERes2Net and WavLM, the training-reward encoder.
Unlike these baselines, its outputs are closer to the target than
the source, with competitive predicted quality but not consistently
the lowest recognition error.
FreeVC \citep{li2023freevc} and Seed-VC receive target-speaker audio
and are evaluated on a separate same-text subset, providing an
upper-bound setting with directly available target timbre.
Our setting instead infers target timbre from source audio and an
editing instruction.
Results with human-recorded ESD references appear in
Appendix~\ref{app:esd_timbre_editing}.
Since target similarity does not establish attribute preservation,
Table~\ref{tab:res_timbre_attrib_op} evaluates requested operations
and unchanged attributes separately.
\textit{EDICT-Stage I} leads in edit direction and categorical accuracy
under both instruction forms, while Qwen3-TTS-VD better preserves
unchanged attributes but follows edits less reliably, highlighting
a trade-off between edit accuracy and preservation.
Up/Down measures direction rather than magnitude;
Appendix~\ref{app:supplementary_results_timbre_editing} details the protocol.
\begin{table}[!t]
\centering
\caption{\textbf{Attribute-operation accuracy (\%) on 500 TimbreEdit-Bench requests.}
Gemini compares source and generated audio without the target recording.
Up/Down assesses ordinal direction, Category assesses categorical targets,
and Preservation assesses unchanged attributes.
Best and second-best scores within each instruction form are
\textbf{bold} and \underline{underlined}.}
\label{tab:res_timbre_attrib_op}
\footnotesize
\setlength{\tabcolsep}{2pt}
\renewcommand{\arraystretch}{1.02}
\resizebox{0.75\linewidth}{!}{%
\begin{tabular}{@{}llcccc@{}}
\toprule
\textbf{System} & \textbf{Instruction}
 & Up Acc.$\uparrow$ & Down Acc.$\uparrow$
 & Category Acc.$\uparrow$ & Preservation$\uparrow$ \\
\midrule
\multirow{2}{*}{Qwen3-TTS-VD}
 & Structured
 & 67.0 & 64.3 & 53.2 & \textbf{83.3} \\
 & Descriptive
 & 60.0 & 63.0 & 49.8 & \textbf{85.0} \\
\addlinespace[1.5pt]
\multirow{2}{*}{Qwen3-TTS-Base}
 & Structured
 & 62.3 & 54.5 & 46.4 & 45.7 \\
 & Descriptive
 & 66.0 & 54.5 & 50.0 & 45.6 \\
\addlinespace[1.5pt]
\multirow{2}{*}{CosyVoice 2}
 & Structured
 & \underline{68.1} & \underline{65.5}
 & \underline{57.7} & \underline{68.2} \\
 & Descriptive
 & \underline{72.2} & \underline{73.3}
 & \underline{69.0} & 66.2 \\
\addlinespace[1.5pt]
\multirow{2}{*}{\textbf{EDICT-Stage I}}
 & Structured
 & \textbf{70.8} & \textbf{80.6}
 & \textbf{89.3} & 66.7 \\
 & Descriptive
 & \textbf{77.9} & \textbf{79.1}
 & \textbf{91.7} & \underline{70.1} \\
\bottomrule
\end{tabular}%
}
\vspace{-12pt}
\end{table}

\subsection{Intra-Utterance Instruction Control}
\label{sec:results_intratts}
Table~\ref{tab:res_intratts_main} exhibits a consistent trend across Qwen3-TTS-VD and CosyVoice 2: the system with individual synthesis and concatenation (\textit{Concat.}) yields the strongest segment-level and emotion accuracy but weaker transitions, while \textit{Joint} favors speaker consistency over local instruction adherence. \textit{EDICT-Stage II} achieves the best transitions and consistently improves SegAcc, EmoAcc, and SIM-I over TED-TTS across both backbones. Despite the different absolute control scores of the two backbones, \textit{EDICT-Stage II } consistently improves SegAcc, EmoAcc, and transition quality over TED-TTS. It indicats that the benefit of KV cache reconstruction extends beyond Qwen3-TTS-VD. This improvement comes with higher WER/CER than the \textit{Joint} setting, but only modest increases over TED-TTS. Speaker consistency remains comparable to TED-TTS, while predicted speech quality changes only slightly. Overall, KV cache reconstruction improves local instruction adherence and transition quality with a modest trade-off in content fidelity.
Appendix~\ref{app:supplementary_results_intratts} reports complementary
MED-TTS results.
\begin{table}[!t]
\centering
\caption{\textbf{Intra-utterance instruction control on IntraTTS-Bench.}
SegAcc, EmoAcc, and Trans measure instruction adherence, emotion
realization, and transition quality in percent. SIM-I measures
inter-segment speaker similarity; WER/CER reports English
WER / Chinese CER in percent. Best and second-best scores within each backbone are
\textbf{bold} and \underline{underlined}, with CER and WER ranked separately.}
\label{tab:res_intratts_main}
\footnotesize
\setlength{\tabcolsep}{2pt}
\renewcommand{\arraystretch}{1.08}
\resizebox{0.9\linewidth}{!}{%
\begin{tabular}{@{}l*{7}{c}@{}}
\toprule
\multirow{2}{*}{\textbf{System}}
    & \multicolumn{3}{c}{\textbf{Control (\%)}}
    & \textbf{Consistency}
    & \multicolumn{3}{c}{\textbf{Content and quality}} \\
\cmidrule(lr){2-4}\cmidrule(lr){5-5}\cmidrule(l){6-8}
    & SegAcc$\uparrow$ & EmoAcc$\uparrow$ & Trans$\uparrow$
    & SIM-I$\uparrow$ & WER/CER (\%)$\downarrow$
    & UTMOS$\uparrow$ & DNSMOS$\uparrow$ \\
\midrule
Qwen3-TTS-VD (Concat.)
    & \textbf{92.44} & \textbf{92.66} & 60.68 & 0.658
    & \underline{4.5} / \textbf{0.9}
    & 3.266 & \underline{3.939} \\
Qwen3-TTS-VD (Joint)
    & 66.80 & 66.37 & 70.66 & \textbf{0.922}
    & \textbf{3.8} / \underline{1.3}
    & \textbf{3.794} & \textbf{3.952} \\
Qwen3-TTS-VD (TED-TTS)
    & 74.25 & 74.02 & \underline{78.03} & 0.911
    & \underline{4.5} / 1.7
    & \underline{3.655} & 3.853 \\
\textbf{Qwen3-TTS-VD (EDICT-Stage II)}
    & \underline{82.30} & \underline{83.60} & \textbf{82.00}
    & \underline{0.915} & 5.1 / 1.8
    & 3.631 & 3.930 \\
\midrule
CosyVoice 2 (Concat.)
    & \textbf{62.43} & \textbf{62.73} & 31.80 & 0.917
    & \underline{6.5} / \underline{3.6}
    & 3.916 & \underline{3.866} \\
CosyVoice 2 (Joint)
    & 49.68 & 50.19 & \underline{69.71} & \textbf{0.939}
    & \textbf{5.4} / \textbf{3.1}
    & 3.877 & \textbf{3.897} \\
CosyVoice 2 (TED-TTS)
    & 58.10 & 58.00 & 63.94 & 0.920
    & 7.5 / 4.6
    & \textbf{4.021} & 3.828 \\
\textbf{CosyVoice 2 (EDICT-Stage II)}
    & \underline{60.56} & \underline{60.10} & \textbf{76.20}
    & \underline{0.924} & 8.1 / 5.5
    & \underline{3.978} & 3.844 \\
\bottomrule
\end{tabular}%
}
\vspace{-12pt}
\end{table}

\subsection{Joint Timbre and Delivery Control}
\label{sec:results_joint_control}
We combine the reference or edited audio from TimbreEdit-Bench with the texts and delivery instructions from IntraTTS-Bench. \textit{Qwen3-TTS-VD (Joint)} and \textit{Qwen3-TTS-VD (Concat.)} take the original reference audio and a global editing instruction as input. \textit{EDICT Stage I + Concat.} and the EDICT pipeline use the same edited reference, followed by either independent synthesis and concatenation or continuous generation with KV cache reconstruction, respectively. 
Table~\ref{tab:res_joint_control} highlights the importance and complementary contributions of the two stages. Compared with \textit{Qwen3-TTS-VD (Concat.)}, incorporating the timbre editor (\textit{EDICT Stage I}) improves target matching, but independently synthesized segments and concatenation (\textit{Concat.}) still exhibit weak transitions in \textit{EDICT Stage I + Concat.}. By further replacing the \textit{Concat.} with KV cache reconstruction, EDICT improves temporal continuity and speaker consistency, albeit at the cost of lower segment-level accuracy and a higher recognition error rate. The EDICT system achieves the highest target similarity and predicted perceptual quality, while maintaining speaker consistency close to that of \textit{Qwen3-TTS-VD (Joint)}. These results demonstrate that an edited reference can effectively guide local delivery changes within a single utterance, although KV cache reconstruction does not fully preserve the segment-level accuracy achieved by independent synthesis.

\begin{table}[!t]
\centering
\caption{\textbf{Joint timbre and delivery control with Qwen3-TTS-VD.}
Requests combine TimbreEdit-Bench voice edits with IntraTTS-Bench
texts and delivery instructions. SIM-T compares the complete output
with the target voice; SIM-I measures inter-segment speaker similarity.
WER/CER reports English WER / Chinese CER in percent.
Best and second-best scores are \textbf{bold} and \underline{underlined},
with CER and WER ranked separately.}
\label{tab:res_joint_control}
\footnotesize
\setlength{\tabcolsep}{2pt}
\renewcommand{\arraystretch}{1.08}
\resizebox{0.9\linewidth}{!}{%
\begin{tabular}{@{}l*{8}{c}@{}}
\toprule
\multirow{2}{*}{\textbf{Configuration}}
    & \multicolumn{2}{c}{\textbf{Control (\%)}}
    & \multicolumn{3}{c}{\textbf{Content and quality}}
    & \textbf{Consistency}
    & \multicolumn{2}{c}{\textbf{SIM-T}$\uparrow$} \\
\cmidrule(lr){2-3}\cmidrule(lr){4-6}
\cmidrule(lr){7-7}\cmidrule(l){8-9}
    & SegAcc$\uparrow$ & Trans$\uparrow$
    & WER/CER (\%)$\downarrow$ & UTMOS$\uparrow$ & DNSMOS$\uparrow$
    & SIM-I$\uparrow$ & WavLM & ERes2Net \\
\midrule
Qwen3-TTS-VD (Joint)
    & 60.19 & \underline{70.07}
    & \textbf{3.6} / 1.2 & 3.339 & 3.869
    & \textbf{0.918} & 0.683 & 0.395 \\
Qwen3-TTS-VD (Concat.)
    & \underline{86.43} & 56.69
    & \underline{4.5} / \textbf{0.8}
    & \underline{3.669} & \underline{3.895}
    & 0.884 & 0.694 & 0.405 \\
\midrule
EDICT Stage I + Concat.
    & \textbf{91.17} & 58.47
    & 4.8 / \underline{1.1} & 3.296 & 3.739
    & 0.881 & \underline{0.749} & \underline{0.598} \\
\textbf{EDICT}
    & 81.62 & \textbf{83.45}
    & 5.1 / 1.9 & \textbf{3.795} & \textbf{3.921}
    & \underline{0.917} & \textbf{0.788} & \textbf{0.609} \\
\bottomrule
\end{tabular}%
}
\vspace{-12pt}
\end{table}

\subsection{Subjective Evaluation}
\label{sec:results_subjective}
\begin{figure}[!t]
    \centering
    \includegraphics[width=0.9\linewidth]{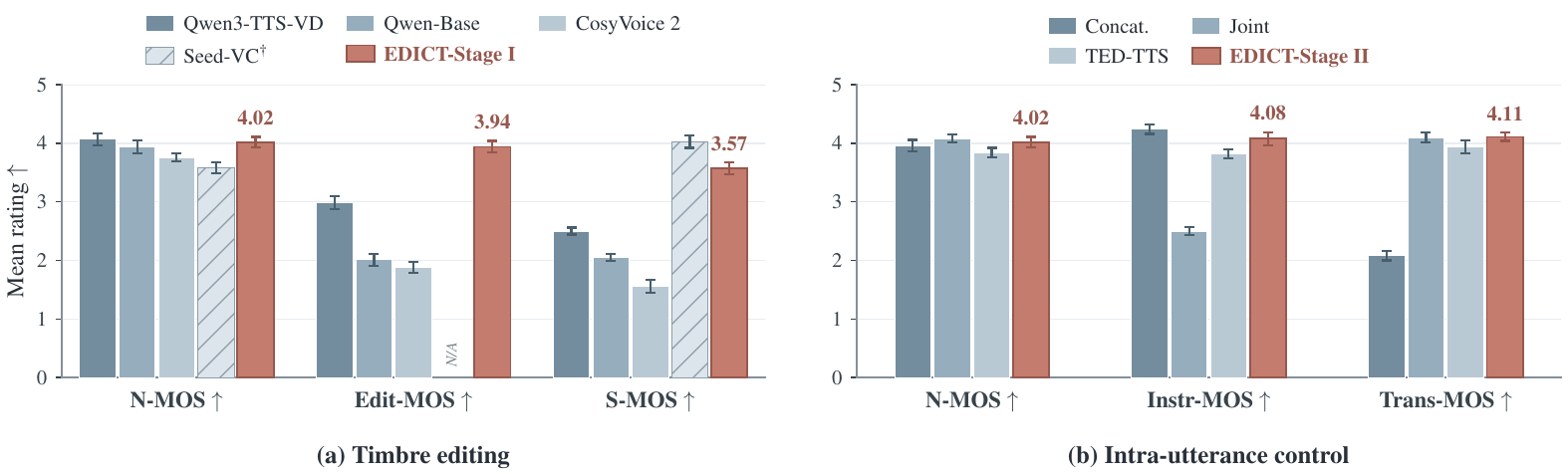}
    \caption{\textbf{Subjective evaluation of timbre editing and local control.}
    Bars show mean five-point ratings from 19 listeners; error bars are
    computed from 95\% bootstrap confidence intervals over audio items.
    Each system has 10 items in (a) and 12 in (b).
    EDICT is highlighted in coral; all variants in (b) use Qwen3-TTS-VD.
    $\dagger$Seed-VC receives target audio and is not rated for
    instruction-based edit accuracy.}
    \label{fig:human_evaluation}
    \vspace{-12pt}
\end{figure}

Figure~\ref{fig:human_evaluation} presents the results of the subjective listening tests. For timbre editing, \textit{EDICT Stage I} achieves the highest mean ratings for edit accuracy and target similarity among instruction-conditioned systems, while maintaining naturalness comparable to Qwen3-TTS-VD. Seed-VC attains higher target similarity, but it is conditioned directly on the target audio rather than a textual instruction. For intra-utterance control, \textit{Concat.} performs best in instruction following but receives the lowest transition-quality rating, highlighting the discontinuities introduced by independent segment synthesis. \textit{EDICT Stage II} outperforms TED-TTS in both instruction following and transition quality, while achieving naturalness and transition quality close to those of \textit{Joint}. These subjective ratings draw same conclusions as in the objective evaluations.


\subsection{Ablation Studies}
\label{sec:results_ablations}
\paragraph{Timbre-editor training.}
Table~\ref{tab:res_timbre_posttraining} shows that more SFT data
improves target matching for structured instructions, but gives
mixed results for descriptive instructions.
A larger candidate pool improves DPO's SIM-T under both instruction
forms. wDPO achieves the highest SIM-T, but yields higher Chinese
CER than SFT (1M), with mixed changes in predicted speech quality.
Appendix~\ref{app:supplementary_results_timbre_editing}
provides further ablations.
\begin{table}[!t]
    \centering
    \caption{\textbf{Timbre-editor training ablations on TimbreEdit-Bench.}
    All post-training variants start from SFT (1M).
    $M$ denotes the candidate count per condition;
    wDPO includes auxiliary SFT as defined in
    Equation~\ref{eq:timbre_weighted_dpo}.
    SIM-T measures target-voice similarity;
    WER/CER reports English WER / Chinese CER in percent.
    Best and second-best scores within each instruction form are
    \textbf{bold} and \underline{underlined},
    with CER and WER ranked separately.}
    \label{tab:res_timbre_posttraining}
    \footnotesize
    \setlength{\tabcolsep}{3pt}
    \renewcommand{\arraystretch}{1.08}
    \resizebox{0.80\linewidth}{!}{%
    \begin{tabular}{@{}ll*{5}{c}@{}}
        \toprule
        \multirow{2}{*}{\textbf{Configuration}}
            & \multirow{2}{*}{\textbf{Instruction}}
            & \multirow{2}{*}{WER/CER (\%)$\downarrow$}
            & \multirow{2}{*}{UTMOS$\uparrow$}
            & \multirow{2}{*}{DNSMOS$\uparrow$}
            & \multicolumn{2}{c}{\textbf{SIM-T}$\uparrow$} \\
        \cmidrule(l){6-7}
            & & & & & WavLM & ERes2Net \\
        \midrule

        \multirow{2}{*}{SFT (0.5M)}
            & Structured
            & \textbf{3.9} / \underline{1.5}
            & 3.689 & 3.895 & 0.736 & 0.498 \\
            & Descriptive
            & 4.2 / 1.5
            & 3.708 & 3.885 & 0.698 & 0.481 \\
        \midrule

        \multirow{2}{*}{SFT (1M)}
            & Structured
            & \underline{4.1} / \textbf{0.9}
            & \textbf{3.936} & 3.896 & 0.757 & 0.561 \\
            & Descriptive
            & \textbf{3.9} / \textbf{0.9}
            & 3.666 & \underline{3.913} & 0.693 & 0.556 \\
        \midrule

        \multirow{2}{*}{DPO ($M=4$)}
            & Structured
            & 5.1 / \underline{1.5}
            & \underline{3.885} & \textbf{3.962}
            & 0.761 & 0.552 \\
            & Descriptive
            & 4.6 / 1.4
            & \underline{3.768} & 3.909 & 0.711 & 0.543 \\
        \midrule

        \multirow{2}{*}{DPO ($M=16$)}
            & Structured
            & 4.5 / 1.7
            & 3.786 & 3.867
            & \underline{0.778} & \underline{0.626} \\
            & Descriptive
            & 4.1 / 1.6
            & 3.724 & 3.906
            & \underline{0.726} & \underline{0.634} \\
        \midrule

        \multirow{2}{*}{\textbf{wDPO} ($M=16$)}
            & Structured
            & \underline{4.1} / 1.7
            & 3.830 & \underline{3.942}
            & \textbf{0.819} & \textbf{0.653} \\
            & Descriptive
            & \underline{4.0} / \underline{1.1}
            & \textbf{3.816} & \textbf{3.963}
            & \textbf{0.793} & \textbf{0.648} \\
        \bottomrule
    \end{tabular}%
    }
\vspace{-12pt}
\end{table}

\paragraph{KV cache reconstruction.}
Table~\ref{tab:res_cache_ablation} compares reconstruction with
retaining or truncating the existing cache. Reconstruction improves
all three control metrics over both alternatives, indicating that
shortening the cache alone does not account for the gains.
Removing initial context improves local accuracy but weakens speaker
consistency; removing recent context sharply degrades transitions.
These complementary effects support using initial context for voice
stability and recent context for continuation. Reconstruction also
produces higher recognition error than the persistent cache, consistent
with the cost in content fidelity seen in the main comparison.
Appendix~\ref{app:supplementary_results_intratts} reports context-length
sweeps.
\begin{table}[!t]
\centering
\caption{\textbf{KV cache reconstruction ablations on IntraTTS-Bench.}
All variants use Qwen3-TTS-VD. The full configuration retains the first
$S=4$ and most recent $K=5$ acoustic positions; the context ablations
remove one of these groups. Persistent and truncated caches reuse
old KV states, whereas EDICT recomputes them.
Best and second-best scores are \textbf{bold} and \underline{underlined},
with WER and CER ranked separately.}
\label{tab:res_cache_ablation}
\footnotesize
\setlength{\tabcolsep}{2pt}
\renewcommand{\arraystretch}{1.08}
\resizebox{0.95\linewidth}{!}{%
\begin{tabular}{@{}l*{7}{c}@{}}
\toprule
\multirow{2}{*}{\textbf{Configuration}}
    & \multicolumn{3}{c}{\textbf{Control (\%)}}
    & \textbf{Consistency}
    & \multicolumn{3}{c}{\textbf{Content and quality}} \\
\cmidrule(lr){2-4}\cmidrule(lr){5-5}\cmidrule(l){6-8}
    & SegAcc$\uparrow$ & EmoAcc$\uparrow$ & Trans$\uparrow$
    & SIM-I$\uparrow$ & WER/CER (\%)$\downarrow$
    & UTMOS$\uparrow$ & DNSMOS$\uparrow$ \\
\midrule
Persistent KV cache
    & 54.25 & 54.02 & 68.00 & \textbf{0.918}
    & \textbf{3.9} / \textbf{0.8}
    & 3.555 & \textbf{3.953} \\
Truncated KV cache
    & 58.05 & 56.70 & 64.94 & 0.892
    & 4.4 / \underline{1.1}
    & 3.278 & 3.905 \\
\midrule
\textbf{EDICT-Stage II}
    & \underline{82.30} & \underline{83.60} & \textbf{82.00}
    & \underline{0.915} & 5.1 / 1.8
    & \textbf{3.631} & \underline{3.930} \\
\quad w/o initial context
    & \textbf{89.29} & \textbf{89.00} & \underline{79.99}
    & 0.863 & 4.3 / 1.4
    & \underline{3.585} & 3.896 \\
\quad w/o recent context
    & 81.15 & 82.26 & 55.32 & 0.875
    & \underline{4.2} / \underline{1.1}
    & 3.548 & 3.830 \\
\bottomrule
\end{tabular}%
}
\vspace{-12pt}
\end{table}

\section{Conclusion}
\label{sec:conclusion}
In this paper, we present EDICT, a framework for combining
reference-conditioned timbre editing with intra-utterance
instruction control in text-to-speech synthesis.
An attribute-based timbre editor produces an edited reference
that anchors the shared voice, while segment-wise KV cache
reconstruction updates local delivery conditions and retains
selected acoustic context.
We introduce TimbreEdit-Bench and IntraTTS-Bench to evaluate
these complementary forms of control.
Objective evaluations, subjective listening tests, and ablation
studies demonstrate improved timbre-edit accuracy and
target-voice matching, alongside a favorable balance between
local instruction adherence, voice consistency, and natural
transitions.
Joint-control experiments further show that edited references can support successive delivery changes within a single utterance.
Future work will expand the timbre-editing data to cover more
diverse voice characteristics and editing requests, and explore
applications in conversational speech generation.
\label{page:main_text_end}

\subsection*{AI use statement}
\phantomsection
\label{sec:ai_use_statement}
We used generative AI tools to assist in constructing synthetic training data and benchmarks, including instruction generation and request parsing, as detailed in \autoref{app:benchmark_construction}. Generative AI was also used in the automatic evaluation of generated speech; the corresponding judge prompts, evaluation protocols, and human verification procedures are described in \autoref{app:evaluation_protocols}. In addition, LLMs were used for language refinement. All generated materials, evaluation procedures, and manuscript content were reviewed and validated by the authors, who take full responsibility for the final content of this work.



\bibliography{references}
\bibliographystyle{iclr2027_conference}

\clearpage
\appendix

\section*{Appendices}
\phantomsection
\label{app:appendices}
This appendix provides the task definitions, data construction,
implementation details, evaluation protocols, and additional analyses.
The contents below link to each section and subsection.

\begingroup
\newcommand{\appendixcontentsentry}[2][0pt]{%
    \par\noindent\hspace*{#1}%
    \hyperref[#2]{\ref*{#2}\quad\nameref*{#2}}%
    \nobreak\dotfill\pageref{#2}\par
}
\setlength{\parskip}{5pt}
\appendixcontentsentry{app:method_comparison}
\appendixcontentsentry{app:training_data}
\appendixcontentsentry{app:benchmark_construction}
\appendixcontentsentry[1.5em]{app:benchmark_construction_timbre_editing}
\appendixcontentsentry[1.5em]{app:benchmark_construction_intratts}
\appendixcontentsentry{app:implementation_details}
\appendixcontentsentry[1.5em]{app:implementation_training}
\appendixcontentsentry[1.5em]{app:implementation_decoding}
\appendixcontentsentry[1.5em]{app:implementation_baselines}
\appendixcontentsentry{app:evaluation_protocols}
\appendixcontentsentry{app:listening_test}
\appendixcontentsentry{app:attention_analysis}
\appendixcontentsentry{app:supplementary_results}
\appendixcontentsentry[1.5em]{app:supplementary_results_timbre_editing}
\appendixcontentsentry[1.5em]{app:supplementary_results_intratts}
\endgroup
\clearpage

\section{Task Taxonomy and Method Comparison}
\label{app:method_comparison}

\begin{table}[!t]
\centering
\caption{\textbf{Taxonomy of speech control tasks.}
Typical control inputs, control targets, and temporal scope are shown.
Inputs are not exhaustive, and tasks may overlap.}
\label{tab:control_taxonomy}
\small
\setlength{\tabcolsep}{4pt}
\renewcommand{\arraystretch}{1.20}
\begin{tabularx}{\linewidth}{@{}l >{\raggedright\arraybackslash}X l l@{}}
\toprule
\textbf{Task} & \textbf{Control input} & \textbf{Control target} & \textbf{Scope} \\
\midrule
Speech content editing & Recording + revised text & Spoken content & Selected span \\
\midrule[0.3pt]
Timbre editing & Reference + edit instruction & Selected timbre attributes & Utterance \\
\midrule[0.3pt]
Voice cloning & Reference speech & Voice identity & Utterance \\
\midrule[0.3pt]
Voice design & Description / attribute specification & Voice characteristics & Utterance \\
\midrule[0.3pt]
Global expressive control & Utterance-level instruction & Emotion, rate, prosody, etc & Utterance \\
\midrule[0.3pt]
Local expressive control & Span-specific instructions & Emotion, rate, prosody, etc & Word / segment \\
\bottomrule
\end{tabularx}
\end{table}

\paragraph{Task distinctions.}
Table~\ref{tab:control_taxonomy} distinguishes six related tasks by
their typical inputs, targets, and temporal scope. Speech content
editing changes the words in an existing recording; timbre editing
changes specified voice attributes while aiming to retain others.
Voice cloning and design specify a voice through an acoustic exemplar
and a description, respectively. 
Global and local expressive control
apply delivery conditions to an utterance or to individual spans. 
Instructions are an interface across these tasks, not a separate task. 

\paragraph{Scope of the capability comparison.}
Table~\ref{tab:control_comparison} records capabilities of the named
models in the cited settings. Its timbre-modification column includes
both source-voice editing and target-voice conversion; a positive entry
does not establish selective preservation of unspecified attributes.
Intra-utterance control concerns delivery assigned to words or segments,
excluding content edits and paralinguistic insertions.
Superscripts identify restricted scope or additional training.
EDICT's voice-cloning and voice-design capabilities are inherited from
its backbone and are not evaluated here; this work evaluates
reference-conditioned editing and segment-level delivery control.
Section~\ref{sec:related_work} discusses the related methods. 
Additional related methods explore several complementary settings. Reference-conditioned synthesis includes
NaturalSpeech~3 \citep{DBLP:conf/icml/JuWS0XYLLST000024},
Seed-TTS \citep{anastassiou2024seed}, 
and F5-TTS \citep{DBLP:conf/acl/ChenN0DWZ0025}.
FC-TTS and Control-TTS separately control timbre and style through
speech references \citep{lee2026fc,lan2025controllable}, 
while \citep{zhao2025prosody, zhao2026comelsinger} supports prosody or pitch control with timbre preservation in
voice conversion and synthesis.
LoopTTS \citep{song2026diagnose} corrects local prosodic defects through diagnosis and
instruction-conditioned re-synthesis.
InstructAudio \citep{qiang2026instructaudio} extends natural-language control to both speech and
music generation.

\begin{table}[!t]
    \centering
    \caption{\textbf{Comparison of speech generation and editing methods.}
    Capabilities refer to the named models in the cited settings.
    Superscripts specify restricted scope or additional training.}
    \label{tab:control_comparison}

    \begingroup
    \small
    \setlength{\tabcolsep}{3pt}
    \renewcommand{\arraystretch}{1.00}
    \renewcommand{\tabularxcolumn}[1]{m{#1}}
    \providecommand{\capyes}{\ding{51}}
    \providecommand{\capno}{\ding{55}}
    \providecommand{\methodentry}[2]{%
        #1\newline{\footnotesize\citep{#2}}%
    }

    \begin{tabularx}{\linewidth}{@{}
        >{\raggedright\arraybackslash}m{0.275\linewidth}
        >{\centering\arraybackslash}m{0.095\linewidth}
        >{\centering\arraybackslash}m{0.075\linewidth}
        >{\centering\arraybackslash}m{0.122\linewidth}
        >{\centering\arraybackslash}m{0.15\linewidth}
        >{\raggedright\arraybackslash}X@{}
    }
        \toprule
        \textbf{Method}
            & \shortstack{\textbf{Zero-shot}\\\textbf{cloning}}
            & \shortstack{\textbf{Voice}\\\textbf{design}}
            & \shortstack{\textbf{Timbre}\\\textbf{modification}}
            & \shortstack{\textbf{Intra-utterance}\\\textbf{control}}
            & \shortstack[l]{\textbf{Control}\\\textbf{formulation}} \\
        \midrule

        \methodentry{VoxInstruct}{DBLP:conf/mm/0002QJZLZ0024}
            & \capyes & \capyes & \capno & \capyes\textsuperscript{a}
            & Instruction-guided synthesis and word-level emphasis \\
        \addlinespace[1pt]

        \methodentry{Qwen3-TTS-Base}{hu2026qwen3}
            & \capyes & \capno & \capno & \capno
            & Reference-based voice cloning \\
        \addlinespace[1pt]

        \methodentry{Qwen3-TTS-VoiceDesign}{hu2026qwen3}
            & \capyes & \capyes & \capno & \capno
            & Description-guided voice design \\
        \addlinespace[1pt]

        \methodentry{Spark-TTS}{wang2025spark}
            & \capyes & \capyes & \capno & \capno
            & Attribute-guided design and reference cloning \\
        \addlinespace[1pt]

        \methodentry{VoiceSculptor}{hu2026voicesculptor}
            & \capyes & \capyes & \capno & \capno
            & Voice design followed by cloning \\
        \addlinespace[1pt]

        \methodentry{CosyVoice 2}{du2024cosyvoice}
            & \capyes & \capno & \capno & \capyes\textsuperscript{b}
            & Instructed cloning and phrase-level tags \\
        \addlinespace[1pt]

        \methodentry{IndexTTS2}{zhou2026indextts2}
            & \capyes & \capno & \capno & \capno
            & Cloning with global emotion and duration control \\
        \addlinespace[1pt]

        \methodentry{FlexiVoice}{chen2026flexivoice}
            & \capyes & \capyes\textsuperscript{c} & \capno & \capno
            & Instructed generation with optional voice reference \\
        \addlinespace[1pt]

        \methodentry{VoxCPM2}{zhou2026voxcpm2}
            & \capyes & \capyes & \capno & \capno
            & Voice design and style-controlled cloning \\
        \midrule

        \methodentry{VoxEditor}{sheng2025voice}
            & \capno & \capno & \capyes & \capno
            & Relative voice-attribute editing \\
        \addlinespace[1pt]

        \methodentry{DreamVC}{DBLP:conf/interspeech/HaiTWQE24}
            & \capno & \capno\textsuperscript{d} & \capyes\textsuperscript{d} & \capno
            & Conversion to a text-described target voice \\
        \addlinespace[1pt]

        \methodentry{Step-Audio-EditX}{yan2025step}
            & \capyes & \capno & \capyes\textsuperscript{e} & \capno
            & Style editing and paralinguistic insertion \\
        \addlinespace[1pt]

        \methodentry{FireRedTTS3-Instruct}{shen2026fireredtts3}
            & \capyes & \capyes & \capyes\textsuperscript{f} & \capno
            & Voice design; acoustic and content editing \\
        \midrule

        \methodentry{WeSCon}{DBLP:conf/nips/WangWGGQMHYWCCW25}
            & \capyes & \capno & \capno & \capyes
            & Word-level emotion and rate control \\
        \addlinespace[1pt]

        \methodentry{TED-TTS}{DBLP:conf/acl/LiangLWLZW26}
            & \capyes & \capno & \capno & \capyes
            & Segment-level emotion and duration control \\
        \midrule

        \rowcolor{gray!10}
        \textbf{EDICT}
            & \capyes\textsuperscript{g} & \capyes\textsuperscript{g} & \capyes & \capyes
            & Reference timbre editing and segment-level delivery \\
        \bottomrule
    \end{tabularx}

    \par\vspace{4pt}
    \begin{minipage}{\linewidth}
        \footnotesize
        \raggedright
        \textbf{Legend.}
        \capyes: reported within the stated scope;
        \capno: not explicitly reported in the cited setting.
        Voice design includes description- or attribute-guided generation
        without input audio.
        Timbre modification includes source-voice editing and conversion.
        Intra-utterance control concerns delivery assigned to words or text spans;
        vocalization insertion and content edits are excluded.
        \par\vspace{2pt}

        \textsuperscript{a}\,Word-level emphasis after additional fine-tuning.
        \par
        \textsuperscript{b}\,Phrase-level emphasis and laughter tags.
        \par
        \textsuperscript{c}\,Description-conditioned generation without a voice reference.
        \par
        \textsuperscript{d}\,DreamVC only, excluding DreamVG; target-voice conversion
        does not establish selective attribute preservation.
        \par
        \textsuperscript{e}\,Voice-related style tags, such as older, child, and girl.
        \par
        \textsuperscript{f}\,Timbre-related editing is limited here to pitch alteration.
        \par
        \textsuperscript{g}\,Cloning and voice design are supported by the EDICT
        system but are not evaluated in this work.
    \end{minipage}
    \endgroup
\end{table}

\section{Timbre-Editing Training Data}
\label{app:training_data}

\paragraph{Attribute schema.}
Table~\ref{tab:timbre_attributes} defines the nine attributes used for
profile generation and edit encoding in
Section~\ref{sec:method_timbre_editing}. Categorical edits specify target
values, while ordinal edits specify signed level changes. Unchanged
attributes receive keep operations.
\begin{table}[!t]
    \centering
    \caption{\textbf{Perceptual attribute schema for timbre editing.}
    Perceived gender and age impression are categorical; the seven
    remaining attributes use ordinal levels 1--3 in the listed order.}
    \label{tab:timbre_attributes}
    \small
    \setlength{\tabcolsep}{5pt}
    \renewcommand{\arraystretch}{1.12}
    \begin{tabularx}{\linewidth}{@{}
        >{\raggedright\arraybackslash}p{0.20\linewidth}
        >{\raggedright\arraybackslash}p{0.28\linewidth}
        >{\raggedright\arraybackslash}X
        @{}}
        \toprule
        Attribute & Values & Perceptual description \\
        \midrule
        Perceived gender
            & Masculine, Feminine
            & Perceived vocal masculinity or femininity. \\
        Age impression
            & Child, Young adult, Mature
            & Perceived age conveyed by the voice. \\
        \midrule
        Pitch register
            & Low, Mid, High
            & Typical pitch range of the voice. \\
        Brightness
            & Dark, Neutral, Bright
            & Perceived prominence of high-frequency energy. \\
        Vocal weight
            & Thin, Medium, Thick
            & Perceived fullness and body of the voice. \\
        Resonance focus
            & Head, Balanced, Chest
            & Resonance perceived as head- or chest-focused. \\
        Roughness
            & Smooth, Moderate, Rough
            & Perceived irregularity in voicing. \\
        Breathiness
            & Clear, Slight, Breathy
            & Audible breath noise during phonation. \\
        Nasality
            & Oral, Slight, Nasal
            & Perceived strength of nasal resonance. \\
        \bottomrule
    \end{tabularx}
\end{table}

\begin{figure}[!t]
    \centering
    \includegraphics[width=0.8\linewidth]{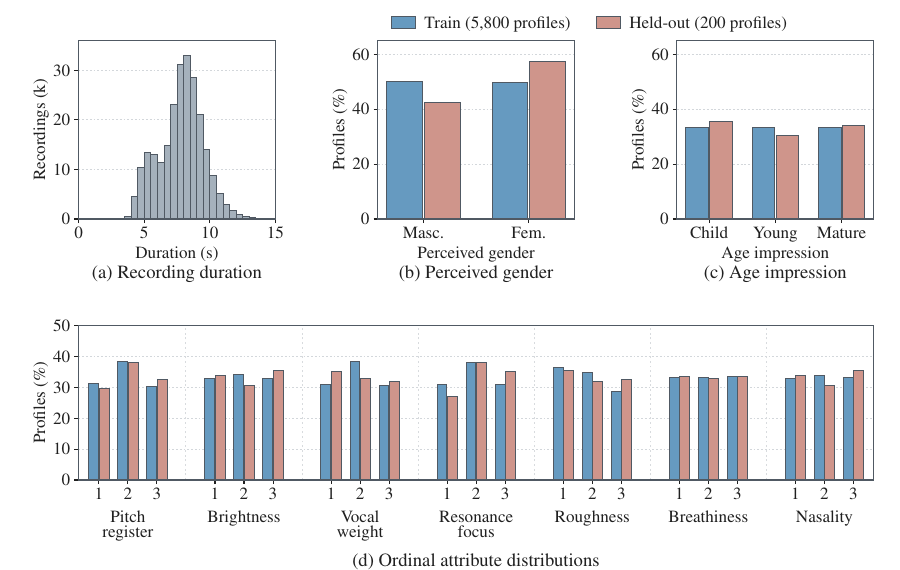}
    \caption{\textbf{Synthetic speech-bank statistics.}
    (a) Retained recording durations in 0.5-second bins.
    (b,c) Perceived gender and age impression.
    (d) Distributions of the seven ordinal attributes; levels 1--3
    follow the schema in Table~\ref{tab:timbre_attributes}.
    Profile percentages are normalized within the training (5,800)
    and held-out (200) subsets. Young denotes young adult.}
    \label{fig:speech_bank_statistics}
\end{figure}

\paragraph{Speech-bank construction.}
SpeechCraft \citep{DBLP:conf/mm/Jin0W0ZZQ024} and CapSpeech \citep{wang2026capspeech} provide speech paired with natural-language
descriptions.
However, these resources primarily describe the characteristics of
individual speech samples, whereas our task requires source--target
pairs specifying which voice attributes to change and which to preserve. 
We therefore construct editing supervision from a synthetic speech bank organized
by the nine-attribute timbre schema introduced in
Section~\ref{sec:method_timbre_editing}.
For each of 6,000 profiles, Qwen3-14B~\citep{yang2025qwen3}\footnote{
\url{https://huggingface.co/Qwen/Qwen3-14B}} produces a natural-language voice
description, which we combine with a shared neutral-delivery instruction
to reduce variation in non-timbre factors. This instruction specifies
a normal speaking rate, medium loudness, and natural pauses, without
whispering, shouting, exaggerated acting, or a regional accent.
The text pool comprises 1,500 Chinese--English parallel pairs.
Each profile is assigned 20 text pairs: five shared across all profiles
and 15 selected from the pool. We synthesize both language versions
with Qwen3-TTS-VD~\citep{hu2026qwen3},\footnote{
\url{https://huggingface.co/Qwen/Qwen3-TTS-12Hz-1.7B-VoiceDesign}}
producing 240,000 recordings.
Filtering is based on generation status and duration; no ASR-based
transcript checks or perceptual attribute screening are applied.
Removing 206 recordings longer than 15 seconds leaves 239,794 recordings
totaling 518.23 hours, with a mean duration of 7.78 seconds.
All durations are derived from codec frame counts at 12\,Hz.
We reserve 200 profiles for constructing TimbreEdit-Bench and use the
remaining 5,800 for training, excluding held-out recordings from both
sides of every training pair.
Figure~\ref{fig:speech_bank_statistics} summarizes the recording durations
and attribute distributions.

\paragraph{Training-pair construction.}
We form directed editing pairs from recordings associated with different
training profiles. The source recording provides the reference voice,
while the target recording and its transcript provide synthesis
supervision. Edit operations are determined by the paired profiles:
signed level differences for ordinal attributes, target categories for
changed categorical attributes, and keep operations for unchanged
attributes. Reusing recordings across pairs yields one million training
examples spanning 826,835 distinct directed profile pairs.
Same-text pairs account for 51.9\% of examples, different-text pairs
within a language for 40.9\%, and cross-language pairs for 7.2\%.
The target language is nearly balanced between Chinese (500,236 examples)
and English (499,764 examples).
Together, the pairs use 229,420 distinct recordings totaling 495.37 hours,
counting each recording once across the source and target sets.
The combined source and target duration averages 13.44 seconds per pair.
Accumulated over all pairs, this amounts to 3,732.51 hours, with reused
recordings counted at each occurrence.

\begin{figure}[!t]
    \centering
    \includegraphics[width=0.8\linewidth]{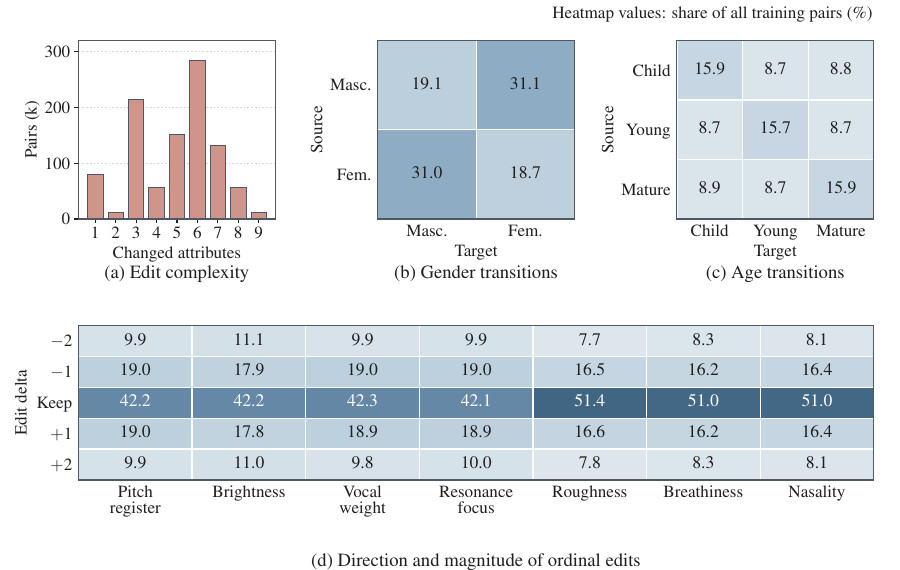}
    \caption{\textbf{Edit coverage across one million training pairs.}
    (a) Number of changed attributes per pair.
    (b,c) Source-to-target transitions for perceived gender and age impression.
    (d) Signed target-minus-source differences for ordinal attributes;
    Keep denotes no change.
    Heatmaps use a shared color scale and show percentages of all pairs.
    Each complete matrix in (b,c) and each column in (d) sums to 100\%
    before rounding. Young denotes young adult.}
    \label{fig:paired_data_statistics}
\end{figure}

\paragraph{Edit coverage.}
The training pairs cover both single-attribute and compositional edits,
including changes of one or two levels in either direction for ordinal
attributes. Of the one million pairs, 80,000 change a single attribute
and 920,000 change multiple attributes, with an average of 4.92 changed
attributes per pair. Every pair specifies at least one change; the
training set contains no all-keep examples.
Figure~\ref{fig:paired_data_statistics} shows the distribution of edit
complexity, categorical transitions, and ordinal edit operations.
These statistics characterize the intended edits derived from profiles;
they do not measure how accurately the synthesized audio realizes them.

\section{Benchmark Construction}
\label{app:benchmark_construction}

Figure~\ref{fig:benchmarks} summarizes the two complementary evaluation tasks.

\subsection{TimbreEdit-Bench}
\label{app:benchmark_construction_timbre_editing}

\paragraph{Pair and instruction construction.}
We construct 2,200 candidate editing pairs from the 200 held-out
timbre profiles described in Section~\ref{sec:experiments_datasets}.
The pairs span nine
categories of single-attribute and compositional edits, including
age, acoustic, and cross-gender changes. Each example provides source
and target recordings, a synthesis text, and two instruction forms.
Structured instructions specify profile-derived attribute changes,
which an LLM rewrites as natural descriptive requests using the
template summarized in Prompt~\ref{prompt:timbreedit_generation}.
Programmatic checks and a separate LLM verifier assess direction,
attribute coverage, intensity, and fluency; rejected candidates are
revised with feedback. Each descriptive variant must preserve every
requested attribute change, its direction, and its target category or
intensity, together with any explicit preservation requirements.
The acceptance rule permits no omitted or added attribute operations,
reversed directions, or altered target categories or intensities,
regardless of edit complexity. Candidates that violate these requirements
are rejected and revised.

\paragraph{Screening and composition.}
Three annotators screen the candidate pairs.
Using the recordings, instruction, and attribute descriptions,
they assess whether the requested changes are audible, unmodified
attributes remain consistent, and audio quality permits a reliable
judgment. The retained benchmark comprises 1,955 pairs: 975 with
Chinese speech and 980 with English speech, including 358
single-attribute and 1,597 compositional edits. Source and target
recordings average 6.40 and 6.38 seconds, respectively.
Prompt~\ref{prompt:timbreedit_generation} is condensed for presentation;
the full template is preserved in the construction materials.

\subsection{IntraTTS-Bench}
\label{app:benchmark_construction_intratts}

\paragraph{Example construction and validation.}
We jointly generate a coherent utterance, two or three consecutive
text segments, and a delivery instruction for each segment,
conditioned on language and topic (dialogue, observation, or vivid
description). Prompt~\ref{prompt:intratts_generation} summarizes the
generation template, which requests natural changes in emotion,
rate, pauses, emphasis, intonation, or energy while excluding timbre
and speaker-identity edits. Programmatic checks validate metadata,
segment counts, nonempty fields, and text lengths, and reject
prohibited identity-related terms. They also require the concatenated
segments to reproduce the full text after whitespace normalization.
These checks are repeated when merging the collection.
Shared voice references are assigned separately during synthesis.

\begin{figure}[!t]
    \centering
    \includegraphics[width=\linewidth]
        {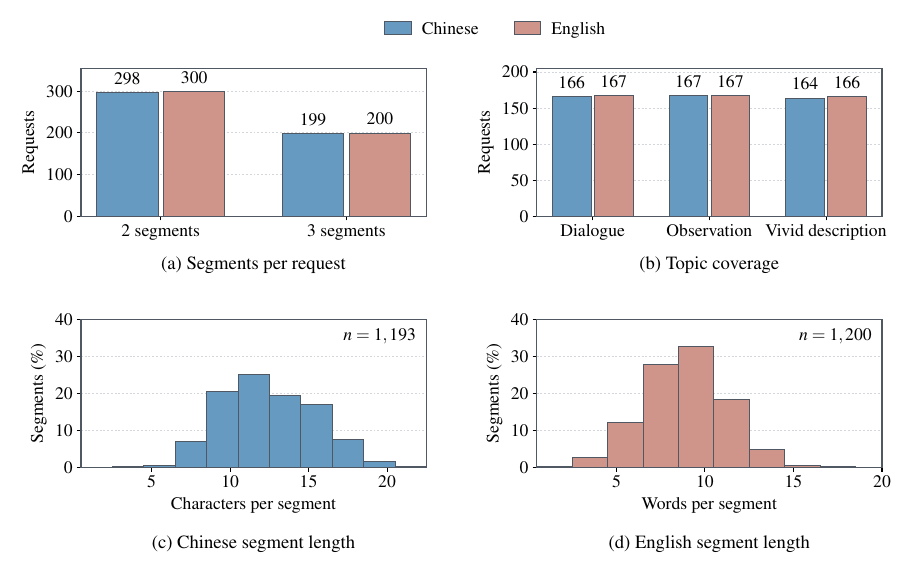}
    \caption{\textbf{IntraTTS-Bench composition.}
    (a,b) Request counts by segment count and topic, split by language.
    (c,d) Segment lengths in bins of width two, normalized within each
    language. Chinese lengths count non-whitespace characters,
    including punctuation; English lengths count
    whitespace-separated words.}
    \label{fig:intratts_statistics}
\end{figure}

\paragraph{Dataset composition.}
IntraTTS-Bench contains 997 requests (497 Chinese and 500 English),
including 598 two-segment and 399 three-segment examples, for a total
of 2,393 segments and 1,396 adjacent boundaries. Dialogue, observation,
and vivid description account for 333, 334, and 330 requests,
respectively. Figure~\ref{fig:intratts_statistics} summarizes the
request composition and segment lengths.
The condensed template in Prompt~\ref{prompt:intratts_generation}
retains the input fields, generation constraints, and output
structure; the complete version is preserved in the construction
materials.

\begin{benchmarkprompt}{TimbreEdit-Bench: descriptive instruction generation}
{prompt:timbreedit_generation}

\textbf{Input.}\par
Source voice description: \texttt{\{ref\_desc\}}\par
Target voice description: \texttt{\{tgt\_desc\}}\par
Structured edit instruction: \texttt{\{canonical\}}

\textbf{Task.}
Rewrite the structured instruction as two natural voice-editing requests
in \texttt{\{language\}}. Use the structured instruction as ground truth:
each variant must independently retain every requested attribute change,
its direction, and its target category or intensity, together with any
explicit preservation requirements. No attribute change may be omitted,
regardless of the number of attributes involved. Do not add, reverse,
strengthen, or weaken any operation. Use the source and target descriptions
only to guide wording, without introducing additional editing requirements.
Describe the desired voice naturally rather than listing attribute
transitions mechanically.
Make clear that the request edits the reference voice. Do not append a
boilerplate statement about keeping other traits unchanged unless it is
explicitly required by the structured instruction. Each variant must be
one sentence, within 80 Chinese characters or 60 English words; shorten
unnecessary wording rather than omit required edits.
Before returning, check each variant against the structured instruction
attribute by attribute and rewrite any variant with missing, added, or
altered operations.

\textbf{Output.} Return JSON only:\par
\texttt{\{"variants": ["...", "..."]\}}
\end{benchmarkprompt}

\begin{benchmarkprompt}{IntraTTS-Bench: segmented text and instruction generation}
{prompt:intratts_generation}

\textbf{Input.}
\texttt{\{language\}}, \texttt{\{topic\}},
\texttt{\{topic\_definition\}}, and \texttt{\{num\_segments\}}.

\textbf{Task.}
Generate one coherent, speakable sentence on the given topic and divide
it into exactly \texttt{\{num\_segments\}} consecutive segments. Assign
one concise delivery instruction to each segment, covering emotion,
rate, pauses, emphasis, intonation, energy, loudness, or speaking manner.
Create clear but natural variation across segments. Do not request
changes to timbre, speaker identity, gender, age, or accent. Avoid rare
named entities, numbers, URLs, code, brackets, and emoji. Target 18--36
Chinese characters or 15--28 English words. Concatenating the segment texts
must reproduce the full sentence with its punctuation and order intact.

\textbf{Output.} Return JSON only; the segment list must have the requested length:\par
{\ttfamily\footnotesize
\{"language": "...", "topic": "...", "num\_segments": N,\par
\hspace*{1em}"text": "...", "segments": [\par
\hspace*{2em}\{"text": "...", "instruction": "..."\}, ...]\}}
\end{benchmarkprompt}

\section{Implementation Details}
\label{app:implementation_details}

\subsection{Timbre-Editor Training}
\label{app:implementation_training}

We initialize the timbre editor from Qwen3-TTS-12Hz-1.7B-Base
and fine-tune all parameters.
SFT runs for 50 epochs on 1M pairs using AdamW with a learning
rate of $4\times10^{-6}$, an effective batch size of 1,024,
weight decay of 0.1, and cosine decay with 10\% warmup.
During initial SFT, the residual-codebook loss has a weight of 0.3,
and conditioning dropout is set to 0.1.
We then perform one round of wDPO with auxiliary SFT.
The frozen SFT model samples $M=16$ candidates for each of 20K
editing conditions drawn from the training pairs, excluding all
200 held-out profiles from both source and target sides.
Post-training uses a constant learning rate of $10^{-6}$,
eight conditions per update, DPO scale $\beta=0.1$,
and auxiliary SFT coefficient $\lambda_{\mathrm{SFT}}=0.01$.
Qwen3-14B parses editing instructions, and a frozen
Qwen3-TTS-VD model performs multi-instruction generation.

\paragraph{Preference-pair construction.}
Preference optimization has been explored for text and speaker
fidelity in TTS \citep{DBLP:conf/emnlp/HussainNYCGFDVL25}
and for non-verbal vocalization synthesis \citep{li2026preference}.
In our method, for each editing condition $g$, we compute a score $s_{g,i}$ for
decoded candidate $i$ as the cosine similarity between its speaker
embedding and that of the paired target recording, using a frozen
WavLM encoder~\citep{chen2022wavlm}.
We select the highest-scoring valid candidate $\mathbf c_g^+$,
with score $s_g^+$, and require $s_g^+\geq0.70$.
An eligible rejected candidate $\mathbf c_{g,j}^-$ must be valid
and satisfy $s_{g,j}^-\geq0.35$ and
$\Delta_{g,j}=s_g^+-s_{g,j}^-\geq0.15$.
We pair each chosen candidate with up to four eligible candidates
having the lowest scores.
Conditions with no eligible pair are excluded.
These preferences are determined by target-speaker similarity and
do not directly account for intelligibility or attribute preservation.

\paragraph{Pair weighting and loss computation.}
Each retained pair is assigned a fixed weight based on its reward gap:
\begin{equation}
    w_{g,j}=\operatorname{clip}\!\left(
        \frac{\Delta_{g,j}}{0.2},\,0.5,\,2.0\right).
    \label{eq:timbre_pair_weight}
\end{equation}
The preference term in Equation~\ref{eq:timbre_weighted_dpo}
is normalized by the sum of pair weights in each minibatch.
We use the frozen SFT sampling model as the reference
$p_{\mathrm{ref}}$.
For both $p_\theta$ and $p_{\mathrm{ref}}$, we evaluate each
chosen and rejected codec sequence under teacher forcing with
its editing condition $g$.
The auxiliary SFT term applies teacher-forced codec cross-entropy
to the chosen sequences under the same conditions, following
prior work on supervised regularization for preference
optimization~\citep{liu2024provably,li2026preference}.
The candidate pool, similarity scores, preference pairs, and
weights remain fixed throughout post-training.

\subsection{Segment-Level Decoding}
\label{app:implementation_decoding}

Generation resumes from the reconstructed cache, which grows
as newly generated acoustic frames are fed back into the model.
Equation 4 describes attention to the rebuilt prefill and the active segment’s acoustic inputs through the current input position.
As illustrated in the upper-right panel of Figure~\ref{fig:kv_recache},
standard causal attention is preserved: subsequent queries
access the reconstructed prefill and the growing segment history.
Intermediate acoustic history excluded during reconstruction
is no longer available in the cache, but its codec frames
remain in the output sequence.
Non-final \texttt{<EOS>} markers trigger reconstruction without
entering the acoustic output, whereas the final
\texttt{<EOS>} terminates generation.
The codec sequences from successive segments form a single
stream that is decoded into the final waveform.
Appendix~\ref{app:attention_analysis} visualizes attention during
cache reconstruction and subsequent decoding.

\subsection{Baseline and Ablation Configurations}
\label{app:implementation_baselines}

\paragraph{Baseline implementations.}
\begin{itemize}[leftmargin=*,itemsep=3pt,topsep=3pt]
    \item \textbf{Timbre editing.}
    Qwen3-TTS-VD and Qwen3-TTS-Base receive the source
    reference through in-context learning, together with the
    editing instruction and target text.
    CosyVoice~2 uses the source recording as its speech prompt
    and receives the edit through its instruction interface.
    We evaluate both structured and descriptive instructions.
    FreeVC and Seed-VC instead receive source and
    target-reference audio without editing instructions,
    and are evaluated separately on a same-text subset.

    \item \textbf{Intra-utterance control.}
    The \textit{Concat.} baseline independently synthesizes each segment
    with its local instruction and directly concatenates the
    waveforms, without crossfading or added silence.
    The \textit{Joint} baseline generates the complete utterance in
    one pass, with an instruction explicitly mapping delivery
    conditions to text spans.
    We adapt TED-TTS to each backbone, using online monotonic alignment
    to switch segment-specific conditions through attention
    masking while retaining causal acoustic history,
    without KV cache reconstruction.

    \item \textbf{Joint control.}
    \textit{Qwen3-TTS-VD (Joint)} and \textit{Qwen3-TTS-VD (Concat.)} receive the original reference, global
    edit request, and local delivery instructions.
    Each \textit{Concat.} request includes the global edit and the
    corresponding local instruction.
    Editor + \textit{EDICT Stage I + Concat.}\ and EDICT use the same precomputed
    edited reference, followed by independent segment
    synthesis or continuous generation with cache
    reconstruction, respectively.
\end{itemize}

\paragraph{Ablation configurations.}
\begin{itemize}[leftmargin=*,itemsep=3pt,topsep=3pt]
    \item \textbf{Timbre-editor training.}
We compare SFT on 0.5M and 1M editing pairs.
Both DPO baselines start from the 1M-pair SFT checkpoint,
use uniform pair weights with auxiliary SFT, and share
the same candidate pool, with $M=4$ subsampled from $M=16$.
DPO-RP ($M=16$) uses the following score for candidate ranking
and margin filtering:
\begin{equation}
    s_i^{\mathrm{RP}}
    = s_i^{\mathrm{tgt}}
    - \lambda\max\!\left(0,s_i^{\mathrm{src}}-\tau\right),
    \qquad
    \tau=0.75,\quad \lambda\in\{0.1,0.2\},
    \label{eq:app_ref_penalty}
\end{equation}
where $s_i^{\mathrm{tgt}}$ and $s_i^{\mathrm{src}}$ are WavLM
speaker-embedding cosine similarities to the target and source,
respectively.
The eligibility thresholds and standard DPO loss remain unchanged. All DPO and wDPO variants include auxiliary SFT. 
Learning-rate and training-budget differences prevent isolating
the effect of the scoring rule.
The additional wDPO round starts from the first-round policy
and uses $M=16$ newly sampled candidates with the target-only reward.

    \item \textbf{KV cache reconstruction.}
    All variants use Qwen3-TTS-VD.
    Persistent KV cache retains the previous acoustic KV
    states across segment switches.
    Truncated KV cache retains only the initial $S$ and
    most recent $K$ acoustic positions but reuses their
    existing KV states.
    EDICT instead reconstructs the cache under the updated
    conditions.
    The full configuration uses $(S,K)=(4,5)$; removing
    initial or recent context gives $(0,5)$ or $(4,0)$,
    respectively, while retaining reconstruction.
\end{itemize}

\section{Automatic Evaluation Protocols}
\label{app:evaluation_protocols}

\paragraph{Objective metrics.}
We transcribe both languages with Whisper large-v3.
Normalization removes punctuation and whitespace for Chinese,
and lowercases and removes punctuation for English.
Utterance-level CER and WER are averaged separately by language.
Speech quality is assessed with UTMOS and DNSMOS (OVRL).
Speaker similarity uses WavLM\footnote{
\url{https://huggingface.co/microsoft/wavlm-base-plus-sv}}
and ERes2Net\footnote{
\url{https://modelscope.cn/models/iic/speech_eres2net_sv_zh-cn_16k-common}}.
SIM-T/R compare complete generated and reference recordings;
joint-control SIM-T uses the paired TimbreEdit-Bench target
recording.
For SIM-I, we use generation-provided segments when available
and forced alignment to segment jointly generated audio.
Similarities are averaged over all unordered segment pairs
within each utterance, then across utterances.

\paragraph{LLM-based evaluation.}
We use Gemini 3.1 Pro\footnote{\url{https://ai.google.dev/gemini-api/docs/models}}.
with temperature zero, without revealing system identities.
For SegAcc and EmoAcc, the judge receives the complete utterance
and its ordered segment texts and instructions, and makes
separate pass/fail judgments of instruction following and
emotion realization. Prompt~\ref{prompt:eval_segment} specifies
these criteria and the required output.
Emotion-eligible segments are identified from the instructions
and held fixed across systems.
A separate call assesses speaker continuity and natural
connection at each boundary, allowing audible but natural
changes in delivery. Prompt~\ref{prompt:eval_transition} gives
the boundary evaluation criteria.
For attribute-operation evaluation, the judge receives source
and generated audio, attribute definitions, and
benchmark-annotated operations, but no target audio.
Prompt~\ref{prompt:eval_attributes} defines separate judgments
for ordinal direction, categorical targets, and explicit keep operations.
For each operation type, pass/fail outcomes are averaged over
applicable attributes within a request, then across eligible
requests.
The condensed templates below summarize the evaluation criteria.

\begin{benchmarkprompt}{Segment-level instruction and emotion evaluation}{prompt:eval_segment}
\textbf{Input.} One complete generated utterance;
\texttt{\{ordered segment texts and instructions\}};
\texttt{\{emotion-eligible segment indices\}}.

\textbf{Task.} Evaluate each segment against its own instruction.
Judge separately whether the delivery matches all explicitly
requested attributes and whether it conveys the requested emotion.
Assess emotion independently of other requirements, such as
speaking rate; use \texttt{not\_applicable} for emotion-ineligible
segments. Do not impose unspecified requirements or use speaker
identity, recording quality, or pronunciation as separate
scoring criteria.

\textbf{Output.} Return JSON with each segment's index, separate
instruction and emotion verdicts (\texttt{pass}/\texttt{fail},
with \texttt{not\_applicable} allowed for emotion), and brief reasons.
\end{benchmarkprompt}

\begin{benchmarkprompt}{Transition evaluation}{prompt:eval_transition}
\textbf{Input.} One complete generated utterance;
\texttt{\{ordered segment texts and instructions\}}.

\textbf{Task.} For each adjacent boundary, judge whether both
sides sound like the same speaker and connect naturally,
without an audible splice, unnatural gap, or abrupt acoustic
reset. Natural pauses and smooth changes in delivery are
acceptable; the boundary need not be imperceptible.
Changes in pitch, energy, or rate do not by themselves indicate
a speaker change. Assess instruction following separately.

\textbf{Output.} Return JSON with the boundary indices, a Boolean
judgment for speaker continuity and natural connection, and a
combined \texttt{pass} when both are satisfied
(\texttt{fail} otherwise), with a brief reason.
\end{benchmarkprompt}

\begin{benchmarkprompt}{Timbre attribute-operation evaluation}{prompt:eval_attributes}
\textbf{Input.} Clip A: source audio; Clip B: generated audio;
\texttt{\{attribute definitions with ordered levels\}};
\texttt{\{required operations\}}.

\textbf{Task.} Evaluate each attribute operation independently.
For \textit{up/down}, compare Clip B with Clip A and judge
the requested direction along the supplied ordered levels,
not exact step size.
For \textit{category}, judge whether Clip B matches the
requested perceived-gender or age category.
For \textit{keep}, judge whether the attribute remains
essentially unchanged, allowing minor natural variation.
Do not use linguistic content, pronunciation, recording
volume, or overall naturalness as evidence of timbre
editing success.

\textbf{Output.} Return JSON with each attribute's name,
required operation, a \texttt{pass}/\texttt{fail} verdict
indicating whether it is satisfied, and a brief reason.
\end{benchmarkprompt}

\paragraph{Human verification of audio-LLM judgments.}
\begin{table}[!t]
\centering
\caption{\textbf{Human--Gemini agreement on intra-utterance control.}
Valid counts refer to segment-level judgments for SegAcc
and EmoAcc, and boundary-level judgments for Trans.
Agreement (\%) uses jointly valid items for each pair.}
\label{tab:agreement_intratts}
\footnotesize
\setlength{\tabcolsep}{5pt}
\renewcommand{\arraystretch}{1.10}
\begin{tabular}{@{}l*{6}{c}@{}}
\toprule
\multirow{2}{*}{\textbf{Criterion}}
    & \multicolumn{3}{c}{\textbf{Valid annotations}}
    & \multicolumn{3}{c}{\textbf{Agreement (\%)}} \\
\cmidrule(lr){2-4}\cmidrule(l){5-7}
    & H1 & H2 & Gemini
    & H1--H2 & H1--Gemini & H2--Gemini \\
\midrule
SegAcc
    & 197 & 198 & 200
    & 90.4 & 83.8 & 85.9 \\
EmoAcc
    & 197 & 198 & 200
    & 85.8 & 83.2 & 85.4 \\
Trans
    & 99 & 100 & 100
    & 91.9 & 93.9 & 96.0 \\
\bottomrule
\end{tabular}
\end{table}

\begin{table}[!t]
\centering
\caption{\textbf{Human--Gemini agreement on timbre editing.}
Each annotator provides 100 valid judgments per criterion.
Agreement (\%) is computed from paired pass/fail labels.}
\label{tab:agreement_timbre}
\footnotesize
\setlength{\tabcolsep}{5pt}
\renewcommand{\arraystretch}{1.10}
\begin{tabular}{@{}l*{6}{c}@{}}
\toprule
\multirow{2}{*}{\textbf{Criterion}}
    & \multicolumn{3}{c}{\textbf{Valid annotations}}
    & \multicolumn{3}{c}{\textbf{Agreement (\%)}} \\
\cmidrule(lr){2-4}\cmidrule(l){5-7}
    & H1 & H2 & Gemini
    & H1--H2 & H1--Gemini & H2--Gemini \\
\midrule
Up
    & 100 & 100 & 100
    & 81.0 & 79.0 & 76.0 \\
Down
    & 100 & 100 & 100
    & 73.0 & 67.0 & 66.0 \\
Category
    & 100 & 100 & 100
    & 82.0 & 84.0 & 78.0 \\
Preservation
    & 100 & 100 & 100
    & 74.0 & 72.0 & 70.0 \\
\bottomrule
\end{tabular}
\end{table}

Two human annotators (H1 and H2) independently assess the
same items using the binary criteria applied to Gemini.
For instruction following and emotion realization, Gemini
provides 200 valid segment-level judgments per criterion,
while H1 and H2 provide 197 and 198, respectively.
For transitions, Gemini and H2 each provide 100 valid
boundary judgments, and H1 provides 99.
Pairwise agreement is computed only over items with valid
labels from both members of the corresponding pair.
As shown in Table~\ref{tab:agreement_intratts},
human--Gemini agreement ranges from 83.2\% to 96.0\%,
with the highest agreement on transition quality.
For timbre editing, all three annotators provide 100 valid
judgments per criterion, covering edit direction,
categorical targets, and attribute preservation.
Table~\ref{tab:agreement_timbre} reports pairwise agreement
for these criteria.
Agreement is lower for downward edits and preservation
in both human--human and human--Gemini comparisons.
This pattern suggests that these fine-grained timbre
judgments are less consistent than the evaluated
intra-utterance criteria.
\section{Subjective Listening Test}
\label{app:listening_test}

\begin{figure}[!t]
    \centering
    \includegraphics[width=\linewidth]{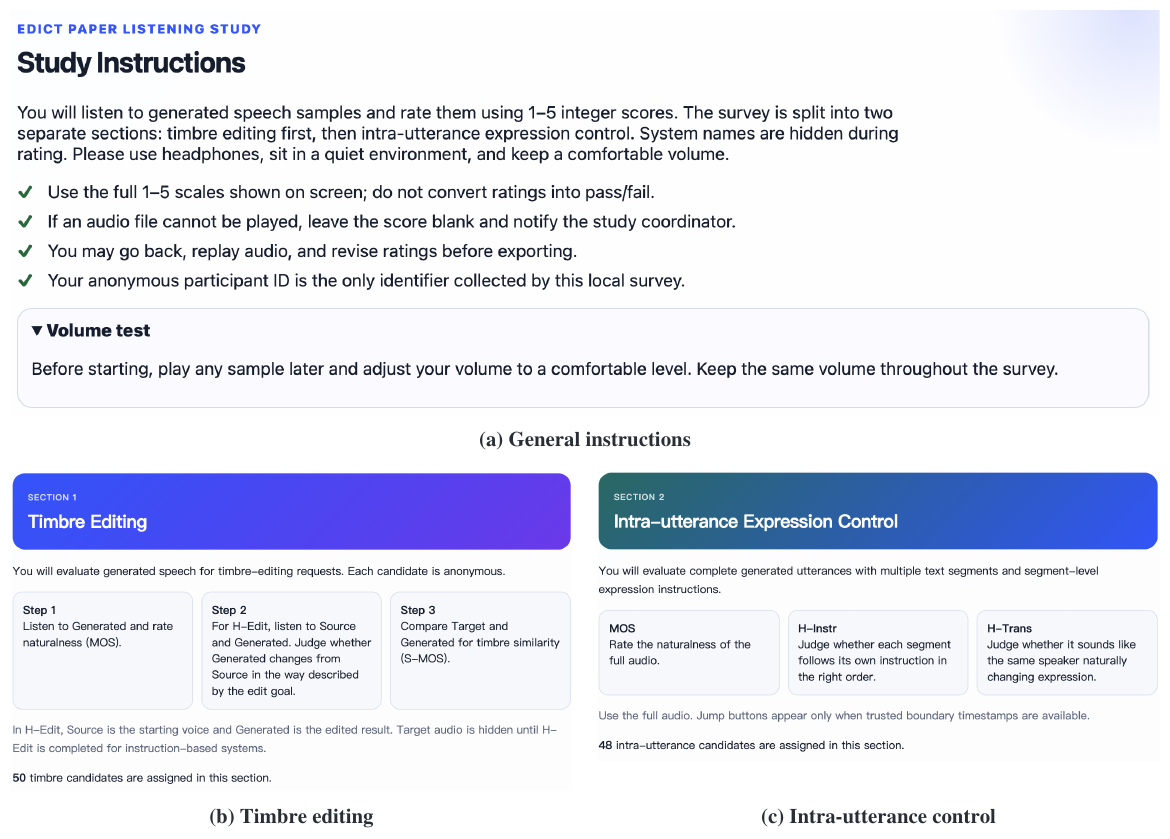}
    \caption{\textbf{Instructions shown to listening-test participants.}
    Panel (a) gives general guidance on playback and rating;
    panels (b) and (c) describe the timbre-editing and intra-utterance
    control tasks, respectively. System identities are hidden.}
    \label{fig:survey_instructions}
\end{figure}

\begin{figure}[!t]
    \centering
    \includegraphics[width=\linewidth]{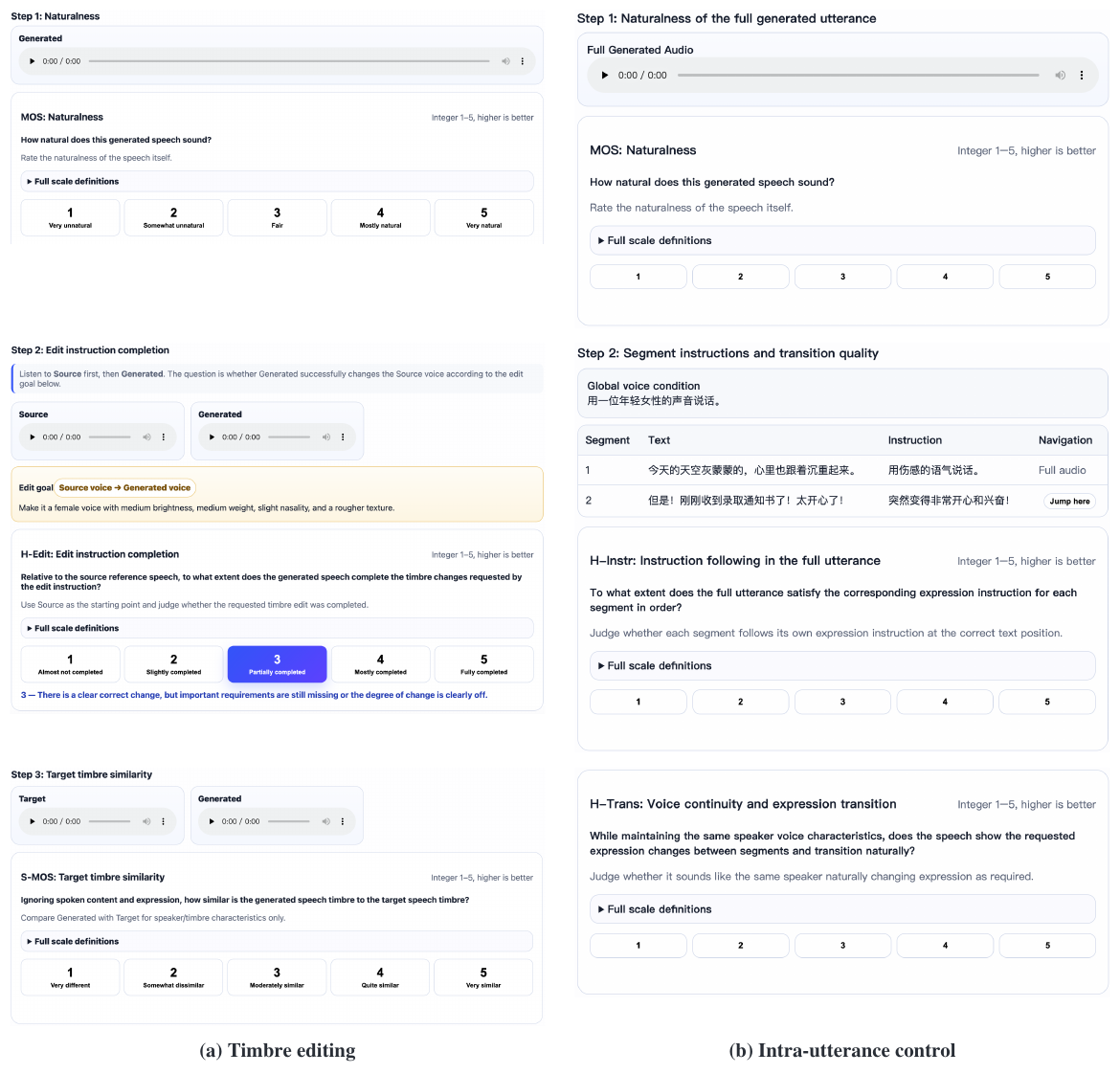}
    \caption{\textbf{Listening-test interfaces.}
    Panel (a) collects naturalness, edit accuracy, and target timbre
    similarity ratings in sequence; the target reference is revealed
    only after edit accuracy is rated.
    Panel (b) collects naturalness, instruction following, and transition
    quality ratings for the complete utterance. All criteria use
    five-point scales.}
    \label{fig:survey_interface}
\end{figure}

\paragraph{Survey procedure.}
The survey covers timbre editing and intra-utterance instruction
control, with system identities hidden from participants.
The study involves 19 participants.
For timbre editing, we evaluate 10 generated audio samples per
system, totaling 50 samples across five systems.
For intra-utterance control, we evaluate 12 samples per system,
totaling 48 samples across four systems.
All criteria use integer ratings from 1 to 5, with higher scores
indicating better performance.
Participants review the general and task-specific guidelines shown
in Figure~\ref{fig:survey_instructions}. They use headphones in a
quiet environment at a comfortable, consistent volume.
They may replay recordings and revise their responses.
No rating is selected by default; playback failures are left unrated.
We report mean item-level listener averages with 95\% percentile
intervals from 10,000 item-bootstrap resamples.
Symmetric error bars use the larger deviation from the mean
to either interval bound. 

\paragraph{Timbre editing.}
Panel (a) of Figure~\ref{fig:survey_interface} shows the three-step
rating interface. Participants first assess generated-speech naturalness,
then compare the source and generated voices against the editing
instruction. The target reference is revealed after edit accuracy
has been rated, allowing target similarity to be assessed separately.
The three criteria are:
\begin{itemize}[leftmargin=*,itemsep=3pt,topsep=3pt]
    \item \textbf{Naturalness (N-MOS).}
    Rate the naturalness of the generated speech, considering
    pronunciation, rhythm, and synthesis artifacts, independently
    of personal voice preference or instruction following.
    The scale ranges from
    \emph{1: very unnatural},
    \emph{2: somewhat unnatural},
    \emph{3: fair},
    \emph{4: mostly natural},
    to \emph{5: very natural}.

    \item \textbf{Edit accuracy (Edit-MOS).}
    Judge how well the generated voice realizes the requested
    changes relative to the source, including explicit preservation
    requirements. Audio quality and target similarity are assessed
    separately.
    The scale ranges from
    \emph{1: almost not completed},
    \emph{2: slightly completed},
    \emph{3: partially completed},
    \emph{4: mostly completed},
    to \emph{5: fully completed}.

    \item \textbf{Target timbre similarity (S-MOS).}
    Compare the generated and target voices, focusing on timbre
    rather than linguistic content, expression, or recording quality.
    The scale ranges from
    \emph{1: very different},
    \emph{2: somewhat dissimilar},
    \emph{3: moderately similar},
    \emph{4: quite similar},
    to \emph{5: very similar}.
\end{itemize}
Target-reference voice conversion is rated for naturalness and
target similarity only.

\paragraph{Intra-utterance instruction control.}
Participants listen to the complete generated utterance while
reviewing its ordered segment texts and delivery instructions.
Panel (b) of Figure~\ref{fig:survey_interface} shows this interface.
Naturalness uses the same N-MOS scale.
Two additional criteria receive one rating each for the full utterance:
\begin{itemize}[leftmargin=*,itemsep=3pt,topsep=3pt]
    \item \textbf{Instruction following (Instr-MOS).}
    Judge whether the requested delivery attributes, such as emotion,
    speaking rate, and prosody, are realized at the corresponding
    text spans. An expression appearing in the wrong segment does
    not fulfill the instruction.
    The scale ranges from
    \emph{1: most instructions unmet or misplaced},
    \emph{2: only a small part followed},
    \emph{3: partially followed with important mismatches},
    \emph{4: mostly followed with minor deviations},
    to \emph{5: all segments follow their instructions}.

    \item \textbf{Transition quality (Trans-MOS).}
    Judge whether the utterance sounds like the same person
    naturally changing expression as instructed.
    High ratings require both voice continuity and appropriate
    expression changes; unchanged delivery alone is insufficient.
    Natural pauses and instruction-appropriate changes in pitch,
    energy, or rate are acceptable, whereas audible splices,
    unnatural gaps, truncation, and apparent speaker changes
    lower the rating.
    The scale ranges from
    \emph{1: severe continuity or transition failures},
    \emph{2: clear problems},
    \emph{3: noticeable flaws},
    \emph{4: natural transitions with minor flaws},
    to \emph{5: smooth, appropriate expression changes with a
    consistent voice}.
\end{itemize}
Throughout the paper, we use N-MOS, Edit-MOS, Instr-MOS, and Trans-MOS.
The original survey interface labels these criteria MOS, H-Edit,
H-Instr, and H-Trans, respectively.

\section{Attention Analysis of KV Cache Reconstruction}
\label{app:attention_analysis}

\begin{figure}[!t]
    \centering
    \includegraphics[width=0.8\linewidth]{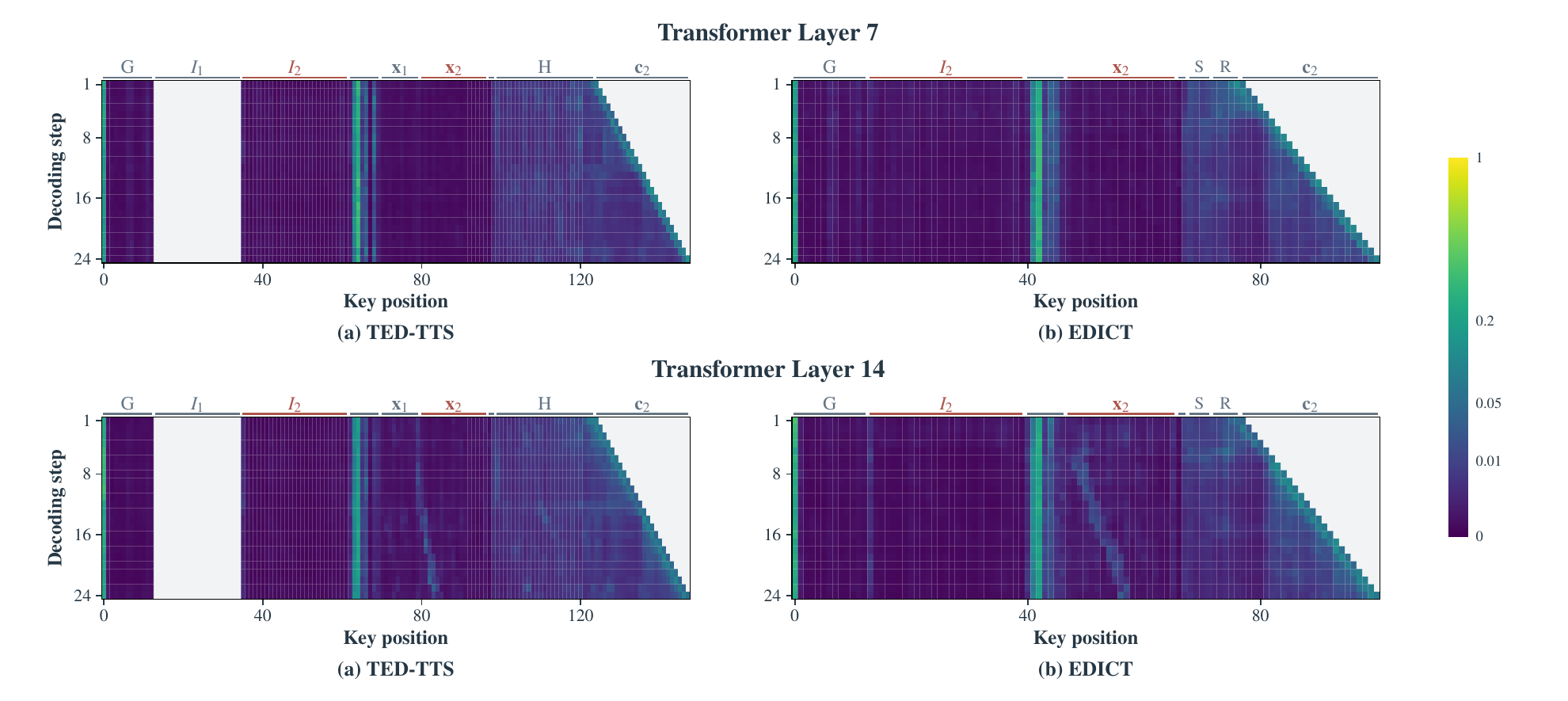}
    \caption{\textbf{Attention across segment transitions.}
    Rows show zero-indexed Transformer layers 7 and 14; columns compare
    TED-TTS and EDICT on the same request, using separate rollouts.
    We average attention over all 16 heads in each layer.
    $G$ denotes the global instruction, $I_k/\mathbf{x}_k$ the segment
    instruction/text, $H$ the preceding audio, $S/R$ the initial/recent
    context, and $\mathbf{c}_2$ the new segment's audio positions.
    Gray cells are masked or unavailable. All panels share a power
    color scale with $\gamma=0.35$.}
    \label{fig:attention_method_comparison_layers}
\end{figure}

\paragraph{Attention across segment transitions.}
Figure~\ref{fig:attention_method_comparison_layers} compares
TED-TTS and EDICT decoding on the same two-segment request.
We visualize Transformer layers 7 and 14 (zero-indexed),
averaging attention weights over all 16 heads within each layer.
In TED-TTS decoding, the previous segment's instruction is
masked, while preceding acoustic states remain in the cache.
EDICT instead constructs a new context from the updated
instruction and text, together with selected initial and recent
acoustic inputs. The maps show subsequent decoding attention
over this reconstructed context and newly generated audio
positions. This distinction concerns both context selection and
representation computation: masking old instructions does not
itself recompute the acoustic states previously conditioned on them.

\begin{figure}[!t]
    \centering
    \includegraphics[width=0.8\linewidth]{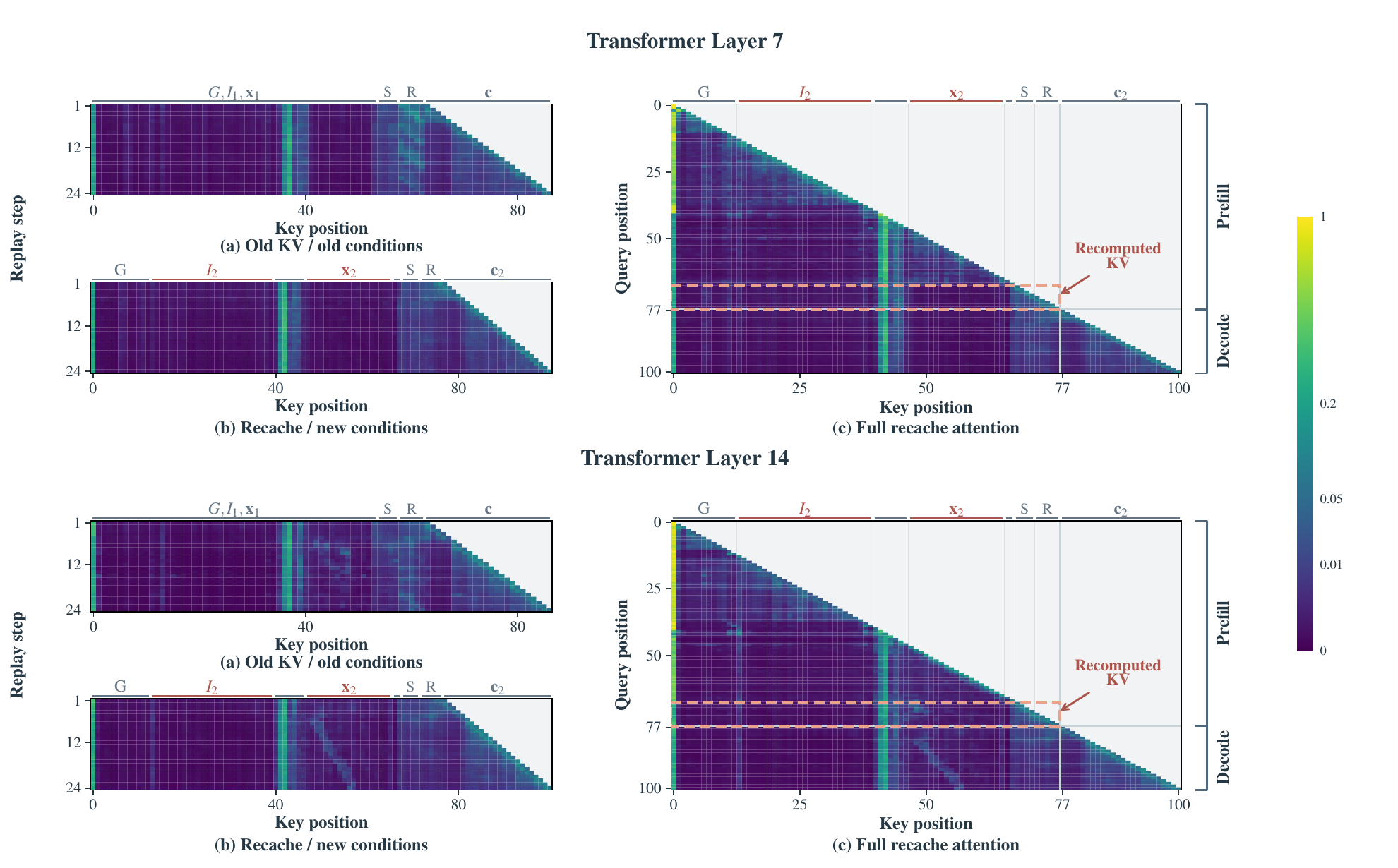}
    \caption{\textbf{Attention during KV cache reconstruction.}
    Each layer group shows replay with (a) old KV states and old
    conditions, (b) reconstructed KV states and new conditions, and
    (c) the full prefill followed by decoding. The dashed coral box
    marks retained acoustic inputs whose KV states are recomputed;
    solid lines separate prefill from decoding.
    Layers 7 and 14 are zero-indexed, with attention averaged over
    all 16 heads. The replay comparison changes both cache and
    conditions, so it does not isolate reconstruction alone.}
    \label{fig:attention_recache_layers}
\end{figure}

\paragraph{Attention during cache reconstruction.}
Figure~\ref{fig:attention_recache_layers} shows the complete
prefill alongside subsequent decoding.
The dashed box marks retained acoustic input positions, whose
representations are recomputed under the new segment conditions
as part of the full prefill. Subsequent decoding attends to the
rebuilt context and newly appended audio positions.
The accompanying replay panels contrast old KV under old
conditions with reconstruction under updated conditions.
This comparison is diagnostic: the old-KV branch receives no
new segment instruction or text, and only codec0 inputs,
rather than all residual codebooks, are fixed across branches.
These visualizations illustrate the computation underlying our
method; the quantitative cache ablations in
Section~\ref{sec:results_ablations} assess its effects on generation
quality and control.

\section{Supplementary Experimental Results}
\label{app:supplementary_results}

This section provides additional experimental results, with audio demonstrations available at \url{https://danny-nus.github.io/EDICT-demopage/}.

\subsection{Timbre Editing with Real Speech Reference}
\label{app:esd_timbre_editing}

Since both our timbre editor’s training data and the reference voices in TimbreEdit-Bench are synthetic, we construct an additional evaluation set from human-recorded ESD dataset \cite{DBLP:journals/speech/ZhouSLL22} to assess whether the learned editing capability extends to real speech references. This set focuses on real human voice compositional edits involving changes to multiple timbre attributes.

\paragraph{Evaluation set construction.}
\begin{figure}[t]
    \centering
    \includegraphics[width=0.85\linewidth]{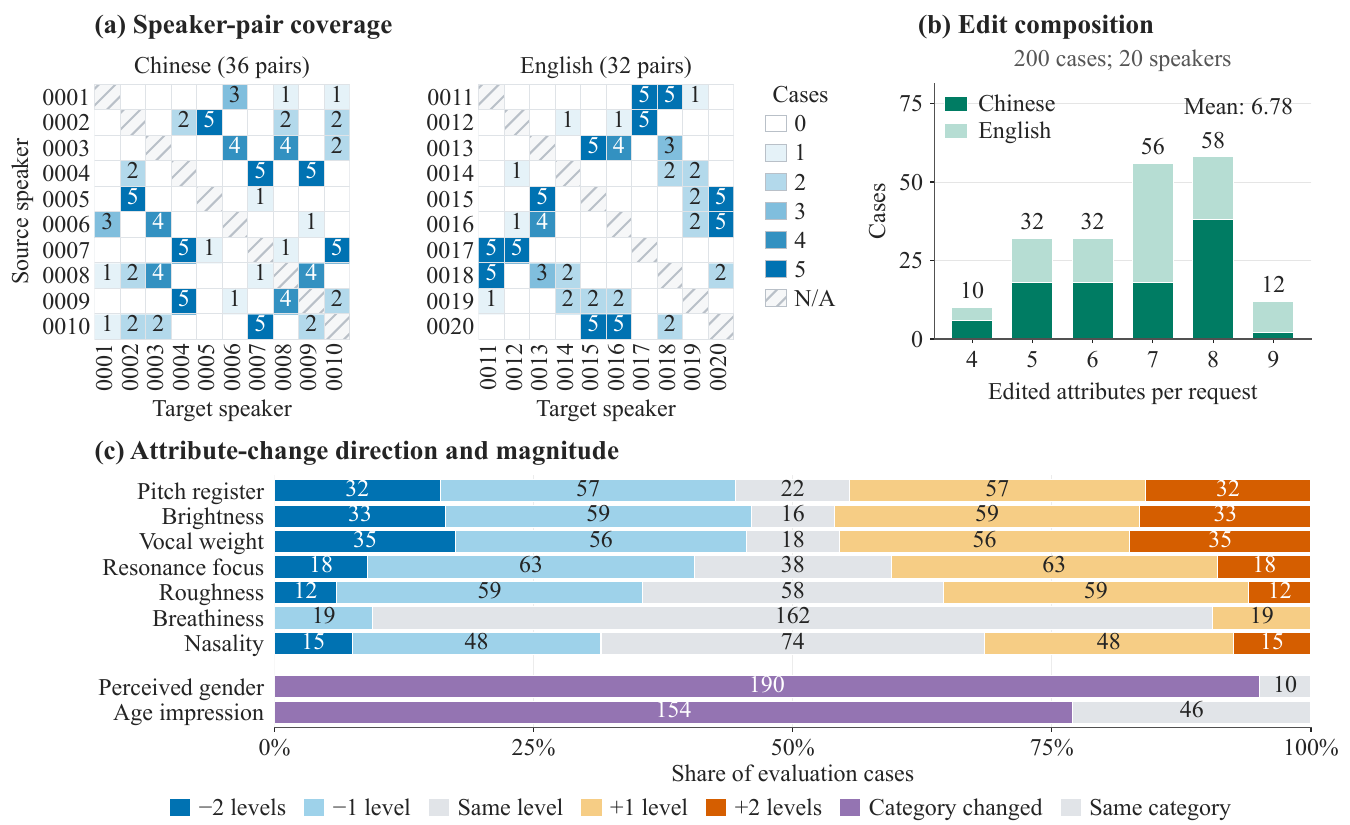}
    \caption{\textbf{Statistics of the ESD evaluation set.}
(a) Speaker-pair coverage.
(b) Number of edited attributes per request.
(c) Attribute-change distributions.
In-bar numbers indicate case counts.}
    \label{fig:esd_real_timbre_statistics}
\end{figure}
We use neutral recordings from 20 Mandarin and English speakers in ESD and manually annotate each speaker using our nine-attribute timbre representation. To construct compositional editing requests, we select speaker pairs with large differences in their annotated timbre attributes. Each case pairs recordings of the same sentence spoken by two different speakers in the same language. Editing requests are derived from the source--target annotation differences, specifying which attributes to modify and which to preserve. The resulting set contains 200 pairs, evenly divided between Mandarin and English, with its statistics information summarized in Figure~\ref{fig:esd_real_timbre_statistics}.

\paragraph{Results and discussion.}
\begin{table}[!t]
\centering
\caption{\textbf{Objective results on the ESD evaluation set.}
WER/CER reports English WER or  Mandarin CER.
SIM-T/R denotes target/source similarity;
$\Delta=\mathrm{SIM\text{-}T}-\mathrm{SIM\text{-}R}$.
Best and second-best scores within each language and instruction form are
\textbf{bold} and \underline{underlined}; SIM-R is unranked.
Target-reference VC is shown separately, with its best
similarity and $\Delta$ scores in \textbf{bold}.}
\label{tab:res_timbre_esd}
\footnotesize
\setlength{\tabcolsep}{2.5pt}
\renewcommand{\arraystretch}{1.24}
\resizebox{0.98\linewidth}{!}{%
\begin{tabular}{@{}ll*{9}{c}@{}}
\toprule
\multirow{2}{*}{\textbf{System}}
 & \multirow{2}{*}{\textbf{Instruction}}
 & \multicolumn{3}{c}{\textbf{Content and quality}}
 & \multicolumn{3}{c}{\textbf{WavLM}}
 & \multicolumn{3}{c}{\textbf{ERes2Net}} \\
\cmidrule(lr){3-5}\cmidrule(lr){6-8}\cmidrule(l){9-11}
 & & WER/CER (\%)$\downarrow$
 & UTMOS$\uparrow$ & DNSMOS$\uparrow$
 & SIM-T$\uparrow$ & SIM-R & $\Delta\uparrow$
 & SIM-T$\uparrow$ & SIM-R & $\Delta\uparrow$ \\
\midrule
\multicolumn{11}{@{}l}{\textbf{(a) English}} \\
\addlinespace[2pt]
\multicolumn{11}{@{}l}{\textit{Instruction-conditioned systems}} \\
\addlinespace[3pt]
\multirow{2}{*}{Qwen3-TTS-VD} & Structured & \underline{5.3} & \underline{4.219} & 3.487 & \underline{0.510} & 0.714 & $\underline{-0.204}$ & \underline{0.383} & 0.531 & $\underline{-0.148}$ \\
 & Descriptive & 6.1 & \underline{4.162} & 3.400 & \underline{0.483} & 0.709 & $\underline{-0.226}$ & \underline{0.385} & 0.533 & $\underline{-0.148}$ \\
\addlinespace[4pt]
\multirow{2}{*}{Qwen3-TTS-Base} & Structured & \textbf{4.7} & \textbf{4.358} & 3.387 & 0.396 & 0.934 & $-0.538$ & 0.301 & 0.770 & $-0.469$ \\
 & Descriptive & \textbf{5.3} & \textbf{4.345} & 3.421 & 0.389 & 0.948 & $-0.559$ & 0.297 & 0.787 & $-0.490$ \\
\addlinespace[4pt]
\multirow{2}{*}{CosyVoice 2} & Structured & 6.5 & 3.992 & \underline{3.607} & 0.446 & 0.792 & $-0.346$ & 0.271 & 0.745 & $-0.474$ \\
 & Descriptive & 8.6 & 4.085 & \textbf{3.595} & 0.467 & 0.809 & $-0.342$ & 0.279 & 0.784 & $-0.505$ \\
\addlinespace[4pt]
\multirow{2}{*}{\textbf{EDICT-Stage I}} & Structured & \underline{5.3} & 4.142 & \textbf{3.639} & \textbf{0.767} & 0.491 & $\textbf{0.276}$ & \textbf{0.594} & 0.330 & $\textbf{0.264}$ \\
 & Descriptive & \underline{5.6} & 4.138 & \underline{3.539} & \textbf{0.728} & 0.486 & $\textbf{0.242}$ & \textbf{0.547} & 0.326 & $\textbf{0.221}$ \\
\midrule
\multicolumn{11}{@{}l}{\textit{Target-reference voice conversion}} \\
\addlinespace[3pt]
FreeVC & N/A & 12.4 & 4.243 & 3.602 & 0.820 & 0.525 & $0.295$ & 0.643 & 0.316 & $0.327$ \\
Seed-VC & N/A & 8.4 & 4.138 & 3.688 & \textbf{0.902} & 0.488 & $\textbf{0.414}$ & \textbf{0.798} & 0.330 & $\textbf{0.468}$ \\
\midrule
\addlinespace[3pt]
\multicolumn{11}{@{}l}{\textbf{(b) Mandarin}} \\
\addlinespace[2pt]
\multicolumn{11}{@{}l}{\textit{Instruction-conditioned systems}} \\
\addlinespace[3pt]
\multirow{2}{*}{Qwen3-TTS-VD} & Structured & \textbf{2.1} & \underline{3.867} & 3.590 & \underline{0.530} & 0.763 & $\underline{-0.233}$ & \underline{0.481} & 0.616 & $\underline{-0.135}$ \\
 & Descriptive & \underline{2.1} & \underline{3.874} & 3.561 & \underline{0.515} & 0.758 & $\underline{-0.243}$ & \underline{0.485} & 0.614 & $\underline{-0.129}$ \\
\addlinespace[4pt]
\multirow{2}{*}{Qwen3-TTS-Base} & Structured & \underline{2.4} & \textbf{3.938} & 3.466 & 0.426 & 0.970 & $-0.544$ & 0.269 & 0.770 & $-0.501$ \\
 & Descriptive & \textbf{1.9} & \textbf{3.956} & 3.428 & 0.411 & 0.968 & $-0.557$ & 0.306 & 0.774 & $-0.468$ \\
\addlinespace[4pt]
\multirow{2}{*}{CosyVoice 2} & Structured & 6.3 & 3.657 & \underline{3.712} & 0.444 & 0.901 & $-0.457$ & 0.244 & 0.742 & $-0.498$ \\
 & Descriptive & 6.7 & 3.701 & \underline{3.723} & 0.421 & 0.939 & $-0.518$ & 0.245 & 0.774 & $-0.529$ \\
\addlinespace[4pt]
\multirow{2}{*}{\textbf{EDICT-Stage I}} & Structured & 3.5 & 3.857 & \textbf{3.726} & \textbf{0.798} & 0.552 & $\textbf{0.246}$ & \textbf{0.583} & 0.269 & $\textbf{0.314}$ \\
 & Descriptive & 3.8 & 3.849 & \textbf{3.727} & \textbf{0.762} & 0.555 & $\textbf{0.207}$ & \textbf{0.536} & 0.271 & $\textbf{0.265}$ \\
\midrule
\multicolumn{11}{@{}l}{\textit{Target-reference voice conversion}} \\
\addlinespace[3pt]
FreeVC & N/A & 2.4 & 3.795 & 3.761 & 0.862 & 0.482 & $0.380$ & 0.631 & 0.293 & $0.338$ \\
Seed-VC & N/A & 1.3 & 3.361 & 3.855 & \textbf{0.947} & 0.475 & $\textbf{0.472}$ & \textbf{0.809} & 0.277 & $\textbf{0.532}$ \\
\bottomrule
\end{tabular}%
}
\end{table}

Table~\ref{tab:res_timbre_esd} shows that \textit{EDICT-Stage I} consistently outperforms the instruction-conditioned baselines in target-speaker similarity across languages and instruction forms, with agreement between the two speaker encoders. Among these systems, only \textit{EDICT-Stage I} achieves positive target--source similarity margins in every condition, indicating that its outputs are, on average, closer to the target than to the source in speaker-embedding space. This advantage persists under descriptive instructions, despite lower target similarity than under structured instructions. \textit{EDICT-Stage I} also achieves the highest DNSMOS in three of the four language--instruction conditions. These results support the transfer of timbre-editing capability from synthetic training data to real speech references on this evaluation set.

\subsection{Timbre Editing with TimbreEdit-Bench}
\label{app:supplementary_results_timbre_editing}

\paragraph{Attribute-level editing and preservation.}
To complement the target-voice similarity results in
Section~\ref{sec:results_timbre_editing}, we evaluate whether
individual attributes change as requested or remain unchanged.
On a fixed subset of 500 TimbreEdit-Bench requests, Gemini
assigns pass/fail judgments to applicable attribute operations.
Up/Down accuracy assesses ordinal edit direction rather than
magnitude, Category accuracy assesses the requested gender
and age categories, and Preservation assesses attributes
marked as unchanged.
Scores are averaged over applicable operations within each
request and then across eligible requests.
As shown in Table~\ref{tab:res_timbre_attrib_op}, \textit{EDICT-Stage I} achieves
the highest Up, Down, and Category accuracy under both
instruction forms.
Qwen3-TTS-VD obtains the highest preservation scores but
lower edit accuracy, highlighting that preservation alone
does not establish successful editing.
\textit{EDICT-Stage I}'s remaining preservation gap indicates that accurately
applying requested changes does not yet guarantee that all
other attributes remain unaffected.

\paragraph{Extended timbre-editor ablations.}
\begin{table}[!t]
\centering
\caption{\textbf{Extended timbre-editor training ablations on TimbreEdit-Bench.}
SIM-T measures target-voice similarity; WER/CER reports 
English WER / Chinese CER in percent.
Best and second-best scores within each instruction form are
\textbf{bold} and \underline{underlined}, with CER and WER ranked separately.}
\label{tab:res_timbre_ablation}
\footnotesize
\setlength{\tabcolsep}{2pt}
\renewcommand{\arraystretch}{1.00}
\resizebox{0.88\linewidth}{!}{%
\begin{tabular}{@{}ll*{5}{c}@{}}
\toprule
\multirow{2}{*}{\textbf{Configuration}}
 & \multirow{2}{*}{\textbf{Instruction}}
 & \multicolumn{3}{c}{\textbf{Content and quality}}
 & \textbf{WavLM} & \textbf{ERes2Net} \\
\cmidrule(lr){3-5}\cmidrule(lr){6-6}\cmidrule(l){7-7}
 & & WER/CER (\%)$\downarrow$
 & UTMOS$\uparrow$ & DNSMOS$\uparrow$
 & SIM-T$\uparrow$ & SIM-T$\uparrow$ \\
\midrule
\multicolumn{7}{@{}l}{\textit{(a) SFT data scale}} \\
\multirow{2}{*}{SFT (0.5M)}
 & Structured
 & \underline{3.9} / 1.5 & 3.689 & 3.895 & 0.736 & 0.498 \\
 & Descriptive
 & 4.2 / 1.5 & 3.708 & 3.885 & 0.698 & 0.481 \\
\addlinespace[1pt]
\multirow{2}{*}{SFT (1M)}
 & Structured
 & 4.1 / \textbf{0.9} & \textbf{3.936} & 3.896 & 0.757 & 0.561 \\
 & Descriptive
 & \underline{3.9} / \textbf{0.9}
 & 3.666 & \underline{3.913} & 0.693 & 0.556 \\
\midrule
\multicolumn{7}{@{}l}{\textit{(b) Post-training from SFT 1M}} \\
\multirow{2}{*}{DPO ($M=4$)}
 & Structured
 & 5.1 / 1.5 & \underline{3.885} & \textbf{3.962} & 0.761 & 0.552 \\
 & Descriptive
 & 4.6 / 1.4 & \underline{3.768} & 3.909 & 0.711 & 0.543 \\
\addlinespace[1pt]
\multirow{2}{*}{DPO ($M=16$)}
 & Structured
 & 4.5 / 1.7 & 3.786 & 3.867 & 0.778 & 0.626 \\
 & Descriptive
 & 4.1 / 1.6 & 3.724 & 3.906 & 0.726 & \underline{0.634} \\
\addlinespace[1pt]
\multirow{2}{*}{\textbf{wDPO} ($M=16$)}
 & Structured
 & 4.1 / 1.7 & 3.830 & \underline{3.942}
 & \textbf{0.819} & \textbf{0.653} \\
 & Descriptive
 & 4.0 / \underline{1.1} & \textbf{3.816} & \textbf{3.963}
 & \underline{0.793} & \textbf{0.648} \\
\midrule
\multicolumn{7}{@{}l}{\textit{(c) Preference-score variants}} \\
\multirow{2}{*}{DPO-RP ($\lambda=0.1$)}
 & Structured
 & 4.6 / 1.9 & 3.764 & 3.880 & 0.798 & 0.631 \\
 & Descriptive
 & 4.1 / 1.8 & 3.625 & 3.891 & 0.746 & 0.614 \\
\addlinespace[1pt]
\multirow{2}{*}{DPO-RP ($\lambda=0.2$)}
 & Structured
 & 4.4 / \underline{1.3} & 3.688 & 3.877 & 0.783 & 0.616 \\
 & Descriptive
 & \underline{3.9} / 1.2 & 3.722 & 3.908 & 0.757 & 0.592 \\
\midrule
\multicolumn{7}{@{}l}{\textit{(d) Additional optimization round}} \\
\multirow{2}{*}{wDPO (round 2)}
 & Structured
 & \textbf{3.8} / 1.5 & 3.754 & 3.897
 & \underline{0.806} & \underline{0.645} \\
 & Descriptive
 & \textbf{3.7} / 1.3 & 3.653 & 3.880 & \textbf{0.812} & 0.631 \\
\bottomrule
\end{tabular}%
}
\end{table}
Table~\ref{tab:res_timbre_ablation} examines SFT data scale,
post-training objectives, preference scoring, and an additional
optimization round.
More SFT data improves target similarity for structured instructions
but gives mixed results for descriptive instructions.
A larger candidate pool improves DPO's target matching, while wDPO
achieves the highest SIM-T among single-round configurations across
both instruction forms and speaker encoders.
Compared with DPO ($M=16$), DPO-RP improves WavLM similarity,
but its ERes2Net results are mixed.
Both RP variants remain below wDPO; differences in pair weighting, learning rate, and training budget prevent attributing this gap solely to the source-similarity penalty.
A second wDPO round reduces English WER and improves descriptive
WavLM similarity, but reduces ERes2Net similarity and predicted
speech quality under both instruction forms.
We therefore retain single-round wDPO as the default to balance
target matching and speech quality.
We examine the auxiliary SFT coefficient below.

\begin{figure}[!t]
    \centering \includegraphics[width=0.9\linewidth]{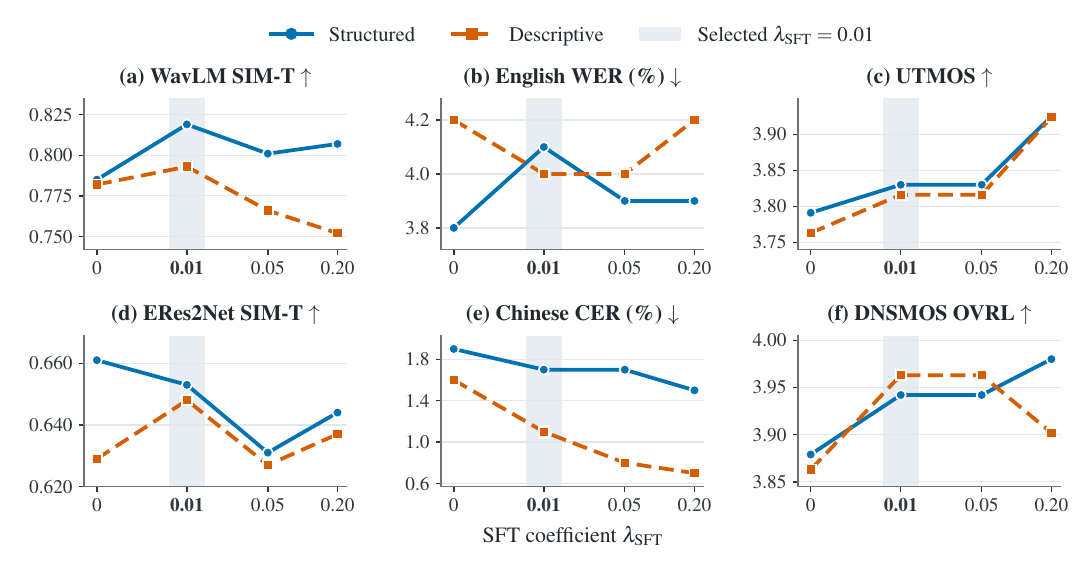}
    \caption{\textbf{Sensitivity to the auxiliary SFT coefficient
    in wDPO on TimbreEdit-Bench.}
    Solid and dashed curves denote structured and descriptive
    instructions, respectively.
    Shading marks the default $\lambda_{\mathrm{SFT}}=0.01$.
    The horizontal axis is linear up to $0.01$ and logarithmic thereafter.}
    \label{fig:sft_weight_sensitivity}
\end{figure}

\paragraph{Effect of auxiliary SFT regularization.}
We vary the auxiliary SFT coefficient
$\lambda_{\mathrm{SFT}}\in\{0,0.01,0.05,0.20\}$ in wDPO;
zero removes the auxiliary SFT term.
As shown in Figure~\ref{fig:sft_weight_sensitivity},
$\lambda_{\mathrm{SFT}}=0.01$ achieves the highest WavLM
target-voice similarity under both instruction forms and the
highest ERes2Net similarity for descriptive instructions.
Larger coefficients can reduce English WER and improve UTMOS,
but do not consistently improve target matching.
We use $\lambda_{\mathrm{SFT}}=0.01$ as the default to prioritize
target-voice similarity for downstream synthesis, accepting
higher recognition error than the lowest-error configurations.

\subsection{Intra-Utterance Instruction Control}
\label{app:supplementary_results_intratts}

\paragraph{Effect of acoustic context length.}

\begin{table}[!t]
    \centering
    \caption{\textbf{Effect of initial acoustic context on a 500-request IntraTTS-Bench subset.}
    With Qwen3-TTS-VD and $K=5$, we vary the number $S$ of initial acoustic
    positions retained during reconstruction.
    SIM-I measures speaker consistency; EmoSim measures emotion-embedding
    similarity to independently synthesized reference segments.
    Bold marks the default configuration, rather than the highest score.}
    \label{tab:context_s}
    \footnotesize
    \setlength{\tabcolsep}{5pt}
    \renewcommand{\arraystretch}{1.05}
    \resizebox{0.6\linewidth}{!}{%
    \begin{tabular}{@{}l*{6}{c}@{}}
        \toprule
        \textbf{Initial context $S$}
            & 0 & 1 & 2 & 3 & \textbf{4} & 5 \\
        \midrule
        SIM-I$\uparrow$
            & 0.872 & 0.885 & 0.902 & 0.899
            & \textbf{0.912} & 0.915 \\
        EmoSim$\uparrow$
            & 0.856 & 0.850 & 0.842 & 0.837
            & \textbf{0.836} & 0.821 \\
        \bottomrule
    \end{tabular}%
    }
\end{table}

\begin{table}[!t]
    \centering
    \caption{\textbf{Effect of recent acoustic context on a 500-request IntraTTS-Bench subset.}
    With Qwen3-TTS-VD and $S=4$, we vary the number $K$ of recent acoustic
    positions retained during reconstruction.
    SIM-I measures speaker consistency; EmoSim measures emotion-embedding
    similarity to independently synthesized reference segments.
    Bold marks the default configuration, rather than the highest score.}
    \label{tab:context_k}
    \footnotesize
    \setlength{\tabcolsep}{3pt}
    \renewcommand{\arraystretch}{1.05}
    \resizebox{0.8\linewidth}{!}{%
    \begin{tabular}{@{}l*{10}{c}@{}}
        \toprule
        \textbf{Recent context $K$}
            & 1 & 2 & 3 & 4 & \textbf{5}
            & 6 & 7 & 8 & 9 & 10 \\
        \midrule
        SIM-I$\uparrow$
            & 0.876 & 0.895 & 0.903 & 0.911
            & \textbf{0.912}
            & 0.912 & 0.915 & 0.917 & 0.916 & 0.912 \\
        EmoSim$\uparrow$
            & 0.837 & 0.842 & 0.838 & 0.832
            & \textbf{0.836}
            & 0.831 & 0.825 & 0.819 & 0.821 & 0.817 \\
        \bottomrule
    \end{tabular}%
    }
\end{table}
Tables~\ref{tab:context_s} and~\ref{tab:context_k} examine
the effects of initial and recent acoustic context, respectively,
on a 500-request subset of IntraTTS-Bench.
We measure speaker consistency using SIM-I and emotional
alignment using EmoSim, the cosine similarity between
emotion2vec embeddings \cite{ma2024emotion2vec} of each generated segment and a
reference segment independently synthesized by Qwen3-TTS-VD
from the same text and local instruction.
Segment scores are averaged within each utterance and then
across utterances.
With $K=5$, increasing $S$ generally improves speaker consistency
but reduces EmoSim. With $S=4$, SIM-I largely plateaus beyond
$K=4$, while larger context windows tend to yield lower EmoSim.
These trends suggest a trade-off between voice consistency
and alignment with the independently synthesized emotional
references, with $(S,K)=(4,5)$ providing a compromise.

\paragraph{Evaluation on MED-TTS.}
\begin{table}[!t]
    \centering
    \caption{\textbf{Intra-utterance emotion control on 500 MED-TTS examples.}
    WER/CER denotes English WER or Chinese CER in percent.
    Emo2v is a segment-level emotion classification accuracy in percent;
    SIM-I measures inter-segment speaker similarity.
    TCD is reported as $10^{-3}\mathrm{TCD}$, and OVRL denotes DNSMOS OVRL.
    Best and second-best scores within each language are
    \textbf{bold} and \underline{underlined}.}
    \label{tab:medtts_overall}
    \footnotesize
    \setlength{\tabcolsep}{3pt}
    \renewcommand{\arraystretch}{1.04}
    \resizebox{0.8\linewidth}{!}{%
    \begin{tabular}{@{}ll*{6}{c}@{}}
        \toprule
        \textbf{Language} & \textbf{System}
            & WER/CER$\downarrow$
            & Emo2v$\uparrow$ & SIM-I$\uparrow$
            & TCD$\downarrow$
            & OVRL$\uparrow$ & UTMOS$\uparrow$ \\
        \midrule
        \multirow{4}{*}{English}
            & Concat.
            & 1.18 & \textbf{88.33} & 0.667
            & 0.467 & \textbf{3.799} & 3.550 \\
            & Joint
            & \textbf{0.30} & 59.47 & \underline{0.929}
            & 0.487
            & \underline{3.754} & \textbf{4.040} \\
            & TED-TTS
            & \underline{0.69} & 80.91 & 0.919
            & \underline{0.463}
            & 3.676 & \underline{3.875} \\
            & \textbf{EDICT-Stage II}
            & 1.09 & \underline{85.81} & \textbf{0.932}
            & \textbf{0.357} & 3.739 & 3.863 \\
        \midrule

        \multirow{4}{*}{Chinese}
            & Concat.
            & \underline{1.74} & \textbf{85.56} & 0.589
            & 0.488 & 3.665 & 3.528 \\
            & Joint
            & \textbf{1.57} & 61.48 & \textbf{0.856}
            & \underline{0.425}
            & \underline{3.692} & \textbf{3.763} \\
            & TED-TTS
            & 2.68 & 76.32 & \underline{0.843}
            & 0.449 & \textbf{3.698} & 3.698 \\
            & \textbf{EDICT-Stage II}
            & 2.47 & \underline{79.96} & 0.838
            & \textbf{0.392}
            & 3.679 & \underline{3.721} \\
        \bottomrule
    \end{tabular}%
    }
\end{table}

\begin{table}[!t]
    \centering
    \caption{\textbf{MED-TTS results by language and text category.}
    WER/CER denotes English WER or Chinese CER in percent.
    Emo2v is a segment-level emotion classification accuracy in percent;
    SIM-I measures inter-segment speaker similarity.
    TCD is reported as $10^{-3}\mathrm{TCD}$, and OVRL denotes DNSMOS OVRL.
    Best and second-best scores within each language and category are
    \textbf{bold} and \underline{underlined}.}
    \label{tab:medtts_categories}
    \footnotesize
    \setlength{\tabcolsep}{3pt}
    \renewcommand{\arraystretch}{1.06}
    \resizebox{0.8\linewidth}{!}{%
    \begin{tabular}{@{}cll*{6}{c}@{}}
        \toprule
        & \textbf{Text category} & \textbf{System}
            & WER/CER$\downarrow$
            & Emo2v$\uparrow$ & SIM-I$\uparrow$
            & TCD$\downarrow$
            & OVRL$\uparrow$ & UTMOS$\uparrow$ \\
        \midrule

        \multirow{12}{*}{\rotatebox[origin=c]{90}{\textbf{English}}}
        & \multirow{4}{*}{\shortstack[l]{Emotional\\Dialogue}}
            & Concat.
            & 2.93 & \textbf{90.18} & 0.659
            & \underline{0.452}
            & \textbf{3.807} & 3.462 \\
        &   & Joint
            & \underline{0.46} & 65.22 & 0.912
            & 0.463
            & \underline{3.708} & \textbf{3.752} \\
        &   & TED-TTS
            & \textbf{0.34} & 84.58 & \underline{0.913}
            & 0.487 & 3.622 & 3.719 \\
        &   & \textbf{EDICT-Stage II}
            & 1.23 & \underline{88.58} & \textbf{0.927}
            & \textbf{0.342}
            & 3.699 & \underline{3.750} \\
        \cmidrule(l){2-9}

        & \multirow{4}{*}{\shortstack[l]{Observational\\Narrative}}
            & Concat.
            & \underline{0.43} & \textbf{84.62} & 0.652
            & \underline{0.469}
            & \textbf{3.820} & 3.525 \\
        &   & Joint
            & \textbf{0.12} & 58.68 & \textbf{0.939}
            & 0.508
            & \underline{3.818} & \textbf{4.197} \\
        &   & TED-TTS
            & 1.00 & \underline{82.69} & 0.921
            & 0.473
            & 3.691 & \underline{3.967} \\
        &   & \textbf{EDICT-Stage II}
            & 1.31 & 81.18 & \underline{0.933}
            & \textbf{0.369}
            & 3.768 & 3.912 \\
        \cmidrule(l){2-9}

        & \multirow{4}{*}{\shortstack[l]{Novelistic\\Description}}
            & Concat.
            & \textbf{0.18} & \textbf{90.19} & 0.690
            & 0.480 & \textbf{3.770} & 3.663 \\
        &   & Joint
            & \underline{0.32} & 54.51 & \textbf{0.936}
            & 0.489 & 3.736 & \textbf{4.171} \\
        &   & TED-TTS
            & 0.73 & 75.46 & \underline{0.923}
            & \underline{0.430}
            & 3.715 & \underline{3.939} \\
        &   & \textbf{EDICT-Stage II}
            & 0.73 & \underline{87.67} & \textbf{0.936}
            & \textbf{0.360}
            & \underline{3.750} & 3.927 \\
        \midrule

        \multirow{12}{*}{\rotatebox[origin=c]{90}{\textbf{Chinese}}}
        & \multirow{4}{*}{\shortstack[l]{Emotional\\Dialogue}}
            & Concat.
            & \underline{1.10} & \textbf{81.85} & 0.586
            & 0.542 & 3.593 & 3.395 \\
        &   & Joint
            & \textbf{0.84} & 61.48 & \textbf{0.827}
            & \underline{0.408}
            & 3.622 & \underline{3.611} \\
        &   & TED-TTS
            & 2.04 & \underline{78.97} & 0.799
            & 0.418 & \textbf{3.656} & 3.559 \\
        &   & \textbf{EDICT-Stage II}
            & 2.03 & 76.26 & \underline{0.808}
            & \textbf{0.399}
            & \underline{3.632} & \textbf{3.627} \\
        \cmidrule(l){2-9}

        & \multirow{4}{*}{\shortstack[l]{Observational\\Narrative}}
            & Concat.
            & \underline{2.26} & \textbf{84.63} & 0.584
            & 0.455 & 3.700 & 3.664 \\
        &   & Joint
            & \textbf{1.87} & 62.41 & \textbf{0.874}
            & \underline{0.435}
            & \textbf{3.744} & \textbf{3.825} \\
        &   & TED-TTS
            & 3.24 & 75.93 & \underline{0.856}
            & 0.462
            & 3.711 & \underline{3.783} \\
        &   & \textbf{EDICT-Stage II}
            & 3.37 & \underline{79.96} & 0.844
            & \textbf{0.417}
            & \underline{3.726} & 3.776 \\
        \cmidrule(l){2-9}

        & \multirow{4}{*}{\shortstack[l]{Novelistic\\Description}}
            & Concat.
            & \textbf{1.86} & \textbf{90.20} & 0.597
            & 0.467 & 3.702 & 3.525 \\
        &   & Joint
            & \underline{2.00} & 60.55 & \underline{0.867}
            & \underline{0.433}
            & \underline{3.710} & \textbf{3.853} \\
        &   & TED-TTS
            & 2.76 & 74.06 & \textbf{0.874}
            & 0.466 & \textbf{3.727} & 3.752 \\
        &   & \textbf{EDICT-Stage II}
            & 2.01 & \underline{83.66} & 0.862
            & \textbf{0.360}
            & 3.679 & \underline{3.760} \\
        \bottomrule
    \end{tabular}%
    }
\end{table}

We extend the evaluation in Section~\ref{sec:results_intratts} to
500 examples from MED-TTS~\citep{DBLP:conf/acl/LiangLWLZW26},\footnote{
\url{https://huggingface.co/datasets/Chanson-0803/MED-TTS}}
using their segment texts and emotion instructions.
Following TED-TTS's text-prompt protocol, \textbf{Emo2v} measures
segment-level emotion classification accuracy with the pretrained
emotion2vec+ large model~\citep{ma2024emotion2vec}.
After ASR alignment, predicted emotions are compared with target
labels over the protocol's five emotion categories.
Emo2v reports percentage accuracy, unlike the embedding similarity
measured by EmoSim. Table~\ref{tab:medtts_overall} compares
\textit{Concat.}, \textit{Joint}, TED-TTS, and EDICT under this protocol.
To additionally assess local acoustic continuity, we report
\textbf{Transition Cepstral Discontinuity} (TCD), a reference-free
boundary metric.
For each internal boundary $b$, we average MFCC vectors within
the 250\,ms windows immediately before and after it and compute
\begin{equation}
    \mathrm{TCD}_b
    = \frac{10\sqrt{2}}{\ln 10}
      \left\|
      \bar{\mathbf{m}}_b^{\mathrm{left}}
      - \bar{\mathbf{m}}_b^{\mathrm{right}}
      \right\|_2,
    \label{eq:tcd}
\end{equation}
where the two vectors represent the mean MFCCs on either side.
We average scores over evaluated boundaries and report
$10^{-3}\mathrm{TCD}$.
Lower values indicate smaller local spectral changes;
this is a continuity proxy rather than a direct measure of
transition naturalness, since phonetic content and intended
expression changes also affect the score.
The aggregate results show the same trade-off as IntraTTS-Bench:
\textit{Concat.}\ provides the strongest emotion control but low speaker
similarity, while \textit{Joint} favors speaker consistency.
\textit{EDICT-Stage II} improves emotion accuracy over TED-TTS in both languages
with comparable speaker similarity. Chinese CER decreases, but
English WER increases. EDICT also obtains the lowest TCD in both
languages. Table~\ref{tab:medtts_categories} shows that its English
emotion gains occur in Emotional Dialogue and Novelistic Description;
TED-TTS remains slightly stronger on Observational Narrative.
\textit{EDICT-Stage II} has the lowest TCD in all three categories of both languages,
indicating smaller boundary spectral changes.


\end{document}